\documentclass[fleqn,usenatbib]{mnras}

\usepackage{newtxtext,newtxmath}

\usepackage[T1]{fontenc}
\usepackage{amsmath}

\usepackage{float} 
\usepackage{placeins}
\DeclareRobustCommand{\VAN}[3]{#2}
\let\VANthebibliography\thebibliography
\def\thebibliography{\DeclareRobustCommand{\VAN}[3]{##3}\VANthebibliography}

\usepackage{graphicx}	
\usepackage{amsmath}	
\usepackage{booktabs}

\usepackage{hyperref} 
\usepackage{url}  

\usepackage{fix-cm}
\usepackage{float}

\title[Characterisation of Hycean hosts]{Characterising M dwarf host stars of two candidate Hycean worlds}

\author[Lalitha Sairam et al.]{
Lalitha Sairam,$^{1}$\thanks{E-mail: lalitha.sairam@ast.cam.ac.uk, nmadhu@ast.cam.ac.uk}
and Nikku Madhusudhan$^{1}$
\\
$^{1}$ Institute of Astronomy, University of Cambridge, Madingley road, Cambridge CB3 0HA, UK\\
}

\date{Accepted March 2025. Received August 2024}

\pubyear{2025}

\begin{document}
\label{firstpage}
\pagerange{\pageref{firstpage}--\pageref{lastpage}}
\maketitle

\begin{abstract}
{Planetary systems orbiting M dwarf host stars are promising targets for atmospheric characterisation of low-mass exoplanets. Accurate characterisation of M dwarf hosts is important for detailed understanding of the planetary properties and physical processes, including potential habitability. Recent studies have identified several candidate Hycean planets orbiting nearby M dwarfs as promising targets in the search for habitability and life on exoplanets. In this study, we characterise two such M dwarf host stars, K2-18 and TOI-732. Using archival photometric and spectroscopic observations, we estimate their effective temperatures (T$_{\mathrm{eff}}$) and metallicities through high-resolution spectral analyses and ages through gyrochronology. We assess the stellar activity of the targets by analysing activity-sensitive chromospheric lines and X-ray luminosities. Additionally, we predict activity cycles based on measured rotation periods and utilise photometric data to estimate the current stellar activity phase. We find K2-18 to be 2.9-3.1 Gyr old with T$_{\mathrm{eff}}$ = 3645$\pm$52 K and metallicity of [Fe/H] = 0.10$\pm$0.12 dex, and TOI-732 to be older (6.7-8.6 Gyr), cooler (3213$\pm$92 K), and more metal-rich ([Fe/H] = 0.22$\pm$0.13 dex). Both stars exhibit relatively low activity making them favourable for atmospheric observations of their planets. The predicted activity cycle and analysis of available high-precision photometry for K2-18 suggest that it might have been near an activity minimum during recent JWST observations, though some residual activity may be expected at such minima. We predict potential activity levels for both targets to aid future observations, and highlight the importance of accurate characterisation of M dwarf host stars for exoplanet characterisation.}

\end{abstract}


\section{Introduction}
The discovery of thousands of exoplanets has sparked intense research into the diversity of their internal structures and atmospheric conditions. Planets orbiting M dwarfs are particularly conducive for atmospheric characterisation. The smaller size and lower brightness of an M dwarf, compared to a sun-like star, yields a higher planet-to-star contrast \citep{Lovis2017, Gilbert-Janizek_2024}. This enables the detection and characterisation of smaller, and potentially habitable, planets through increased transit frequencies and transit depths\citep{Dressing2013, SHIELDS20161}. Due to their lower luminosities, M dwarfs also have habitable zones located closer to the star compared to Sun-like stars, increasing the likelihood of finding Earth-like and Hycean worlds within them \citep[e.g.][]{Owen2016, Kopparapu_2013, Madhusudhan_2021}. However, robust characterisation of planetary atmospheres and accurate assessment of habitability around M dwarfs require a comprehensive understanding of stellar properties, particularly activity levels.

Stellar activity can significantly impact the habitability of a planet, especially for terrestrial-size exoplanets \citep{Owen2016}. M dwarfs are notorious for frequent flares and coronal mass ejections that can strip away the planetary atmospheres and potentially render them unsuitable for life as we know it \citep{Estrela2020, Amaral_2022}. However, some studies suggest that habitability might still be possible, even under these seemingly challenging conditions \citep{Lobo_2023}. Stellar activity also plays a role in shaping planetary systems during their formation \citep{Lueftinger2020}. Studying stellar activity levels in M dwarfs can offer insights into the processes influencing planetary formation and evolution around these stars, aiding in understanding the diversity of exoplanetary systems \citep{Ribas2007}.

Atmospheric characterisation of exoplanets orbiting M dwarfs also face a challenge due to possible stellar contamination \citep{Rackham2018}. This refers to the presence of heterogeneities on the stellar surface, such as spots (darker regions) and faculae/plages (brighter regions), that can significantly impact transmission and emission spectral observations. As the star rotates, these features can induce variations in brightness and spectral signatures, and can influence the observed spectra of exoplanets orbiting such stars (\citealt{Rackham2023}). 

Stellar features like spots and plages can impact transit depth measurements, hindering the precise determination of exoplanet radii (\citealt{Apai2018}).
These features cause variations in the out-of-transit flux level, potentially masking or mimicking features in exoplanetary transmission spectra \citep{Oshagh_2014}. The distinct temperature profiles of spots and faculae cause them to emit different spectra compared to the stellar photosphere.
This can contaminate the observed spectrum of the exoplanet, hindering the accurate retrieval of its atmospheric composition (\citealt{Pinhas2018, Boldt_2020,Thompson_2024}). Given these challenges, robust characterisation of the fundamental properties of M dwarf stars is crucial for effectively interpreting data on exoplanetary atmospheres, particularly those obtained with high-precision instruments on large facilities such as the James Webb Space Telescope (JWST). 

JWST observations of temperate exoplanets orbiting M dwarfs provide a promising avenue in the search for habitable conditions and biosignatures in exoplanets. In recent years, a new class of habitable planets called "Hycean" planets have emerged as exciting targets in this direction \citep{Madhusudhan_2021}. These planets are characterised by ocean covered surfaces with H$_2$-rich atmospheres, with potential habitable conditions in the oceans. Several candidate Hycean planets orbiting M dwarfs have been identified in recent studies making them important targets for atmospheric characterisation with JWST \citep{Madhusudhan_2021, Savvas2022}. The recent discovery of carbon-bearing molecules in the candidate Hycean world K2-18~b is a promising step in this direction \citep{Madhusudhan2023}. 

Despite the growing interest in these systems, a comprehensive characterisation of their M dwarf host stars remains relatively unexplored.  Previous studies have primarily focused on estimating stellar parameters such as effective temperature (T$_{\mathrm{eff}}$), metallicity, and rotation period in some cases. However, a detailed assessment of stellar activity levels, including factors like chromospheric emission, X-ray luminosity, and potential activity cycles, is important for interpreting spectroscopic observations of the planets. Enhanced stellar activity can introduce significant contamination into the planetary spectra, potentially mimicking atmospheric features or masking them altogether \citep{Rackham2023}. Additionally, a robust age estimation for these stars is valuable for modelling atmospheric evolution, as the stellar age can provide insights into factors like the stellar wind and chromospheric activity levels over time \citep{Ribas2007, Vidotto_2014}.

In the present study, we focus on two such M dwarf hosts, K2-18 and TOI-732, both of which have been identified to host candidate Hycean worlds \citep{Madhusudhan_2021}. We conduct a detailed characterisation of both targets, focusing particularly on their activity levels and age estimation. We use a combination of photometric and spectroscopic archival observations and employ established analysis techniques to investigate various aspects of stellar activity, including chromospheric emission, X-ray luminosity, and long-term photometric variations. We further utilise a gyrochronology approach based on stellar rotation periods and empirical relations from open stellar clusters to estimate the ages of both stars. Our findings provide important inputs for both spectroscopic observations and theoretical modelling of planetary conditions in these systems.

In Section \ref{sec:targets_obs}, we introduce the basic properties of the host stars, and describe the spectroscopic and photometric observations used in this work. In Section \ref{sec:analysis}, we report our analyses and results for both the targets, followed by the summary and discussion in section \ref{sec:summary}.
\section{Target Systems and Observations}\label{sec:targets_obs}
In this study we focus on two M dwarf stars, K2-18 and TOI-732, which are known to host temperate sub-Neptunes, including candidate Hycean worlds. Here, we present an overview of the currently known information regarding the host stars and the planets orbiting them. A summary of the stellar properties of both targets from previous studies is presented in Table~\ref{tab:stellar_planetary_parameters}. We also discuss the various spectroscopic and photometric datasets used in this work to characterise these targets.

\begin{table}
\caption{System properties for K2-18 and TOI-732, as reported in the literature.}
\label{tab:stellar_planetary_parameters}
\begin{tabular}{llll}
\hline
\textbf{Parameter} & \textbf{K2-18} & \textbf{TOI-732} \\
\hline
\multicolumn{3}{c}{\textbf{Host Star}} \\
Spectral Type & M2.5$^{(1)}$& M3.5 $^{(6)}$\\
Distance (pc) & 38.099$\pm$0.038$^{(2)}$ &22.027$\pm$0.014$^{(2)}$\\
V (mag) & 13.477$\pm$0.042$^{(3)}$ & 13.140$\pm$0.035$^{(3)}$\\
T$_{\mathrm {eff}}$ (K) & 3590$\pm$93$^{(4)}$ &3360$\pm$51$^{(6)}$\\
R$_{\star}$ (R$_{\mathrm{\odot}}$) & 0.450$\pm$0.013$^{(4)}$ &0.380$\pm$0.012$^{(8)}$\\
Activity & Moderate$^{(4)}$ & Quiet$^{(7)}$\\
\hline
\multicolumn{3}{c}{\textbf{Hycean candidate}} \\
Planet Name & K2-18b &TOI-732c\\
P$_{\mathrm{orb}}$ (days) & 32.93962$\pm$1.0$\times10^{-4}$$^{(5)}$& 12.25228$\pm$1.3$\times10^{-5}$$^{(8)}$\\
Semi-major axis, a$_{\mathrm{p}}$ (au) &0.1591$\pm$0.0004$^{(5)}$ &0.0757$\pm$0.0018$^{(8)}$\\
R$_{\mathrm{pl}}$ (R$_{\oplus}$) & 2.51$^{+0.13}_{-0.18}$$^{(4)}$ & 2.39$^{+0.10}_{-0.11}$$^{(8)}$\\
M$_{\mathrm{pl}}$ (M$_{\oplus}$) & 8.63$\pm$1.35$^{(5)}$ & 8.04$^{+0.50}_{-0.48}$$^{(8)}$\\
\hline

\end{tabular}

\textbf{Notes:} (1) \citealt{Benneke2017}, (2) \citealt{Gaia2022}, (3) \citealt{Zacharias2012}, (4) \citealt{Hardegree-Ullman2020}, 
(5) \citealt{Cloutier2019}, (6) \citealt{Nowak2020},  (7) \citealt{Cloutier_2020} (8) \citealt{Bonfanti2024}  
\end{table}

\subsection{K2-18}\label{sec:k2-18}

K2-18, a red dwarf star in Leo Minor at a distance of 38.099$\pm$0.038 pc \citep{Gaia2022}, stands out as a promising candidate in the search for life due to the possibility of hosting a Hycean world  \citep{Madhusudhan_2021}. The exoplanet K2-18~b, initially reported by \cite{Foreman-Mackey_2015} based on transit observations with the K2 mission, was confirmed by \cite{Montet_2015} and  subsequent RV observations \citep{Cloutier_2017,Sarkis_2018}. K2-18~b orbits its host star with an orbital period of $\sim$32.9 days and a separation of 0.16 au \citep{Cloutier2019}. 
The planet receives an insolation close to that received by the Earth, at 0.92-1.08$\times$ Earth value, placing it in the terrestrial habitable zone \citep{Benneke2019}. Considering a stellar mass of 0.44$\pm$0.01 M$_{\odot}$ \citep{Hardegree-Ullman2020}, the inner edge of the nominal terrestrial habitable zone lies at 0.16 au \citep{Kopparapu_2013} and that for Hycean planets lies below 0.1 au \citep{Madhusudhan_2021}. 
The planet has a radius of 2.51$^{+0.13}_{-0.18}$ R$_{\oplus}$ \citep{Hardegree-Ullman2020} and a mass of 8.63$\pm$1.35 M$_{\oplus}$ \citep{Cloutier2019}, resulting in a density of 3.00$^{+0.66}_{-0.88}$ g cm$^{-3}$. Given its relatively low density, K2-18b is inconsistent with a purely rocky interior and is expected to possess a volatile-rich interior, with a significant component of H$_2$O and/or H$_2$ \citep{Cloutier_2017, Benneke2017, Madhusudhan2020}. Initial atmospheric observations of the planet suggested the presence of an H$_2$-rich atmosphere \citep{Tsiaras2019, Benneke2019, Madhusudhan2020}, albeit with significant degeneracies \citep{Blain2021, Barclay2021}. Furthermore, the extent of the H$_2$-rich envelope is unknown, allowing for a degenerate set of internal structures that can explain the bulk density \citep{Madhusudhan2020}.

Recent JWST observations of K2-18b revealed the presence of methane (CH$_4$) and carbon dioxide (CO$_2$) in a H$_2$-rich atmosphere, marking the first detection of carbon-bearing molecules on a habitable-zone exoplanet \citep{Madhusudhan2023}. This discovery strengthens the case for classifying K2-18b as a Hycean world \citep{Madhusudhan_2021}. Notably, the observed high abundances of CO$_2$ and CH$_4$, and the dearth of detectable NH$_3$ and CO, are consistent with predictions for an ocean surface under a thin H$_2$-rich atmosphere as expected for a Hycean world \citep{Hu2021, Tsai2021, Madhusudhan_2023a}. Furthermore, the composition is inconsistent with predictions for other internal structures, such as a mini-Neptune or gas dwarf, both of which require a deep H$_2$-rich atmosphere \citep{glein_geochemical_2024, Cooke2024, Rigby_towards}. The JWST observations also provided tentative hints of the possible presence of dimethyl sulfide (DMS), a molecule associated with biological activity on Earth, although further confirmation is needed \citep{Madhusudhan2023}. 

\subsection{TOI-732}\label{sec:toi732}
The TOI-732 system, discovered in 2020, offers a unique glimpse into the diversity of exoplanets orbiting M-dwarf stars. Located approximately 22.027$\pm$0.014 pc (\citealt{Gaia2022}) from Earth, TOI-732 is an M3.5 dwarf host star with an effective temperature of approximately 3360$\pm$51 K (\citealt{Nowak2020}). This relatively inactive star provides a potentially stable environment for its orbiting planets, reducing the likelihood of threats to habitability posed by stellar flares and high-energy radiation.

The TOI-732 system comprises of at least two planets (\citealt{Cloutier_2020,Nowak2020}). Follow-up observations with ground-based telescopes and the CHaracterising ExOPlanet Satellite \citep[CHEOPS;][]{benz2021} further confirmed their existence and provided additional timeseries photometry (\citealt{Bonfanti2024}). The first planet, TOI-732 b, falls into the category of ultra-short-period planets, with an orbital period of 0.76837931$\pm$0.0000004 days \citep{Bonfanti2024}.  With a radius of 1.325$^{+0.057}_{-0.058}$ R$_{\oplus}$ and a mass of  2.46$\pm0.19$ M$_{\oplus}$ \citep{Bonfanti2024}, TOI-732 b presents a fascinating example of a compact and dense exoplanet. The second planet in the system, TOI-732 c, orbits its star at a more moderate distance with a period of roughly 12.25228$\pm$0.00001 days \citep{Bonfanti2024}. Larger than its counterpart, TOI-732 c has a radius of 2.39$^{+0.10}_{-0.11}$ R$_{\oplus}$ and a mass of  8.04$^{+0.50}_{-0.48}$ M$_{\oplus}$ (\citealt{Bonfanti2024}). Notably, this planet is considered a potential Hycean candidate, meaning it may possess a thin H$_2$/He atmosphere and a surface ocean, with conditions conducive to microbial life (\citealt{Madhusudhan_2021}). The two planets in this system occupy opposite sides of the M-dwarf radius valley, a region where the occurrence rate of planets dips significantly \citep{Fulton2017}. This positioning makes the TOI-732 system particularly valuable for studying how exoplanets form and evolve under different conditions. Additionally, both planets, particularly TOI-732 c with its larger size and potential Hycean nature, offer promising targets for atmospheric characterisation studies using large telescopes like the JWST.

\subsection{Spectroscopic data}

Table~\ref{tab:obs} summarises the spectroscopic observations used in this study to analyse the K2-18 and TOI-732 systems. We utilised high-resolution spectroscopic data obtained using several ground-based facilities, as follows:
\begin{itemize}
    \item HARPS (High Accuracy Radial velocity Planetary Searcher): Mounted on the 3.6-meter ESO telescope at La Silla Observatory, Chile, HARPS boasts an exceptional resolving power (R~$\sim$110,000). HARPS observes a wide optical wavelength range (378-691 nm), allowing for detailed analysis of stellar properties through absorption lines \citep{Mayor2003}. HARPS provide high-stability data with a typical internal precision of less than 1 ms$^{-1}$ for radial velocity measurements, ideal for detecting potential exoplanets. Publicly available HARPS spectra, including those used in this study, can be queried and downloaded from the ESO archive using the generic query \texttt{form3} \footnote{\href{https://archive.eso.org/wdb/wdb/adp/phase3_main/form}{https://archive.eso.org/wdb/wdb/adp/phase3\_main/form}}. This form provides access to all phase 3 data reduced by the HARPS data-reduction software (DRS\footnote{\href{https://www.eso.org/sci/facilities/lasilla/instruments/harps/doc/DRS.pdf}{https://www.eso.org/sci/facilities/lasilla/instruments/harps/doc/DRS.pdf}}).  In this study, HARPS provided high-quality spectra for 107 epochs of K2-18 spread over 1190 days and 33 epochs of TOI-732 observed over 247 days.
    \item CARMENES: Located at the Calar Alto Observatory in Spain, CARMENES is another high-resolution spectrograph (R~$\sim$85,000) dedicated to exoplanet research. Compared to HARPS, CARMENES observes in the near-infrared regime (0.52-1.7 $\mu$m). CARMENES boasts a stability of around 2 ms$^{-1}$ for radial velocity measurements \citep{Quirrenbach_2010}. Publicly available CARMENES spectra including all reduced spectra and associated products for the M dwarfs observed during guaranteed-time observations can be found in \cite{Ribas_2023}
\footnote{\href{https://carmenes.caha.es/ext/pressreleases/DR1/}{https://carmenes.caha.es/ext/pressreleases/DR1/}}.
    CARMENES data for K2-18 covers 59 epochs over 431 days, while for TOI-732, it spans 52 epochs over a period of 54 days.
    \item Immersion Grating Infrared Spectrometer (IGRINS): Mounted on the McDonald Observatory's 2.7-meter Harlan J. Smith Telescope, IGRINS is a high-resolution spectrograph (R~$\sim$45,000) specifically designed for the near-infrared region covering wavelength range 1.5-4.5 $\mu$m  (\citealt{Yuk_2010, Jeong_2014, Park2014}). Unlike HARPS and CARMENES, IGRINS focuses on the crucial H and K bands within the near-infrared, which can provide valuable insights into the atmospheres of stars and potentially even exoplanets. IGRINS obtained 27 epochs of TOI-732 data over a concentrated period of 2.8 hours. The IGRINS Pipeline Package (PLP; \citealt{Lee2017}) was used by the instrument team for data reduction, spectral extraction, and wavelength calibration. A description of the reduced data used in this work can be found in \cite{Cabot_2024} and \cite{Cheverall_2024}.

\end{itemize}

\begin{table}
\caption{Number of epochs of spectroscopic and photometric observations of K2-18 and TOI-732}
\label{tab:obs}
\begin{tabular}{lcc}
\hline
{\textbf{Instrument}}&{\textbf{K2-18}}&{\textbf{TOI-732}}\\
&{\textbf{Epochs (timespan)}}&{\textbf{Epochs (timespan)}}\\
\hline
{\textbf{Spectroscopy}}&&\\
HARPS &107 (1190-d) & 33 (247-d)\\
CARMENES&59 (431-d) & 52 (54-d)\\
IGRINS&--&27 (2.8-h)\\
{\textbf{Photometry}}&&\\
ASAS&328 (2575-d)&46 (1508-d)\\
CATALINA&362 (2726-d)&306 (1258-d)\\
MEarth&--&24372 (1889-d)\\
Kepler &3554 (80-d)&--\\
STELLA&27 (108-d) &--\\
\hline

\end{tabular}

\end{table}

\subsection{Photometric data}

This study utilises archival photometric data from several ground- and space-based telescopes.

\begin{itemize}
    \item The All-Sky Automated Survey for Supernovae (ASAS\footnote{\href{https://www.astrouw.edu.pl/asas/?page=main}{All-Sky Automated Survey}}; \citealt{Pojmanski_2000}) is a ground-based photometric CCD sky survey that has been continuously monitoring the entire southern sky and a portion of the northern sky ($\delta < 25^{\circ}$) since October 2000. The ASAS system uses four instruments, each offering distinct capabilities: Two wide-field cameras, boasting a field of view of 9$^{\circ} \times 9^{\circ}$, capture data through B and V filters. Additionally, a single very wide-field camera, spanning an impressive 36 $^{\circ} \times 36^{\circ}$, observes through a R filter. This diverse set of instruments allows ASAS to effectively monitor large areas of the sky while simultaneously capturing variability information across a broad range of stellar types. The ASAS observations provide nearly continuous coverage for K2-18, spanning 2575 days with 328 data points (epochs). For TOI-732, ASAS data offers a less frequent sampling with 46 epochs collected over 1508 days.
    \item The CATALINA Sky Survey (CATALINA\footnote{\href{http://nunuku.caltech.edu/cgi-bin/getcssconedb_release_img.cgi}{The CATALINA} Survey}; \citealt{Christensen_2012}) leverages a network of telescopes strategically positioned in the Santa Catalina Mountains north of Tucson, Arizona. This ground-based system prioritises wide-area sky surveillance, particularly useful for near-Earth object (NEO) discovery and long-term stellar variability studies. For this study, CATALINA contributes photometric data for both K2-18 and TOI-732. The K2-18 data offers a comprehensive view, spanning 2726 days with 362 epochs, suggesting a slightly higher observation frequency compared to ASAS. However, TOI-732 observations from CATALINA are similar to ASAS, with 306 epochs collected over 1258 days.
    \item Launched in 2009, the Kepler space telescope was a NASA mission specifically designed to detect exoplanets using the transit method (\citealt{Borucki_2010}). This spacecraft employed a single, large-aperture (1.5-meter) telescope equipped with a high-precision photometer. The archival data from the Kepler spacecraft  covers the 3554 epochs over 80 days for K2-18 \footnote{\href{https://archive.stsci.edu/k2/preview.php?dsn=KTWO201912552-C01&type=LC}{K2 data search and retrieval for K2-18}}. 
    \item High-precision optical photometry for K2-18 was obtained with STELLA, a twin 1.2-meter robotic telescope observatory in Tenerife, Spain (\citealt{Strassmeier_2004}). STELLA prioritises high-precision photometric measurements for cool stars, offering a combination of a large aperture and efficient scheduling for maximised data collection. 
    This study utilises STELLA data spanning 108 days with 27 epochs as from \cite{Sarkis_2018}, providing high-precision optical photometry for K2-18 .
    
    \item The MEarth Project\footnote{\href{https://lweb.cfa.harvard.edu/MEarth/DataDR11.html}{The MEarth project}}, dedicated to finding exoplanets orbiting cool M dwarf stars, utilises two robotic observatories (\citealt{Nutzman_2008}). Located in both hemispheres (Arizona and Chile), these observatories house eight identical 0.4-meter telescopes each. Equipped with infrared-sensitive CCD cameras, MEarth continuously monitors target stars, searching for transiting exoplanets. This automated system allows for efficient observation of a large number of stars. MEarth contributes extensive data for TOI-732, spanning 24372 epochs over 1889 days.
\end{itemize}

\begin{figure}
\includegraphics[width=0.49\textwidth]{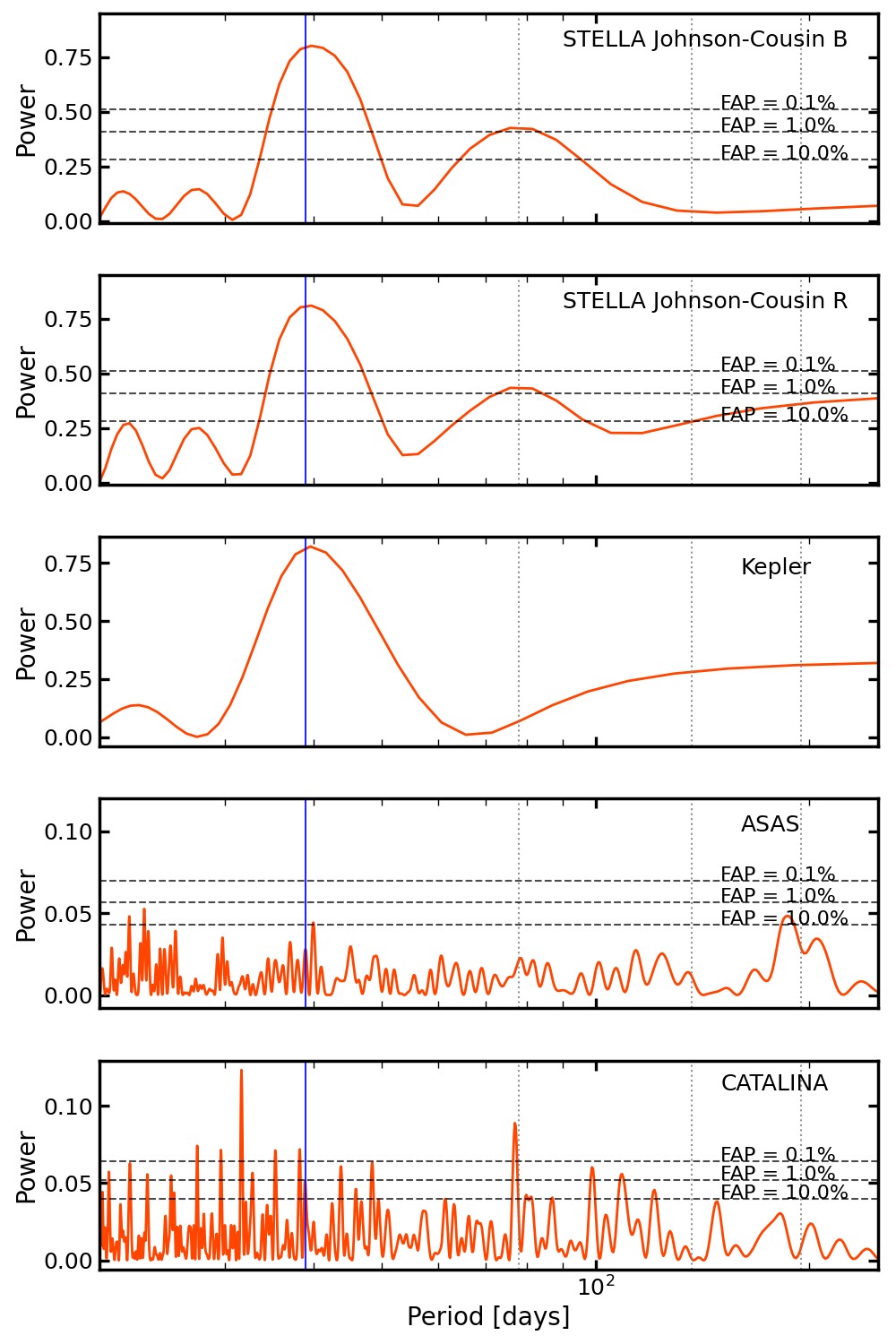}
\caption{The periodogram for K2-18 is presented for STELLA, Kepler, ASAS and CATALINA data. The vertical blue lines indicates the stellar rotation period (around 39 days) and its harmonics (the dotted lines). The three horizontal dashed lines across the plot correspond to false alarm probabilities of $10\%, 1\%$, and $0.1\%$. Note that for the Kepler data (3rd panel), the FAP levels are not visible as they are almost overlapped with the x-axis due to the extremely strong signal.}

\label{fig:rotationperiod_periodogram}
\end{figure}

\begin{figure}
\includegraphics[width=0.49\textwidth]{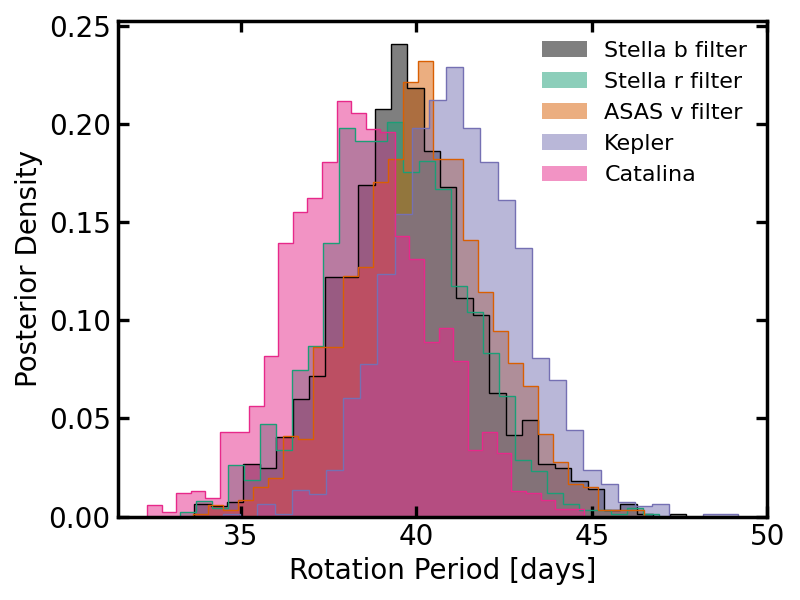}
\includegraphics[width=0.49\textwidth]{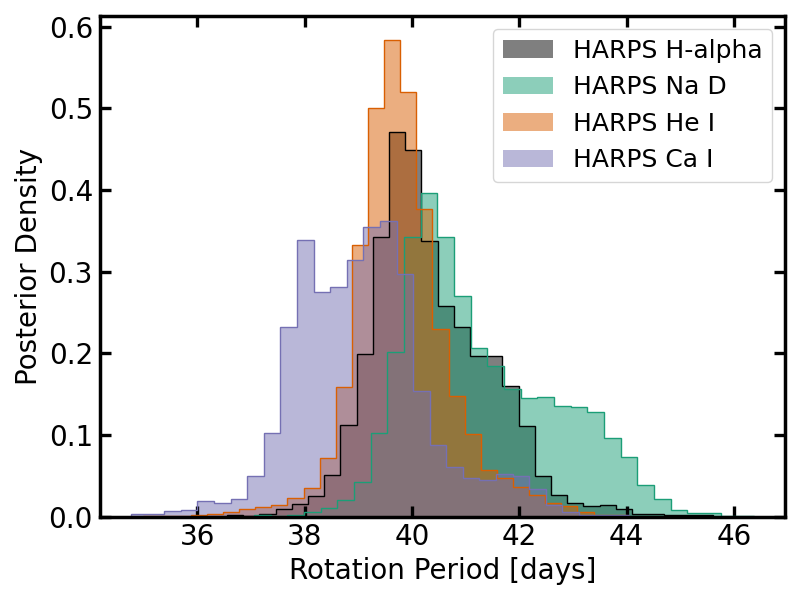}
\includegraphics[width=0.49\textwidth]{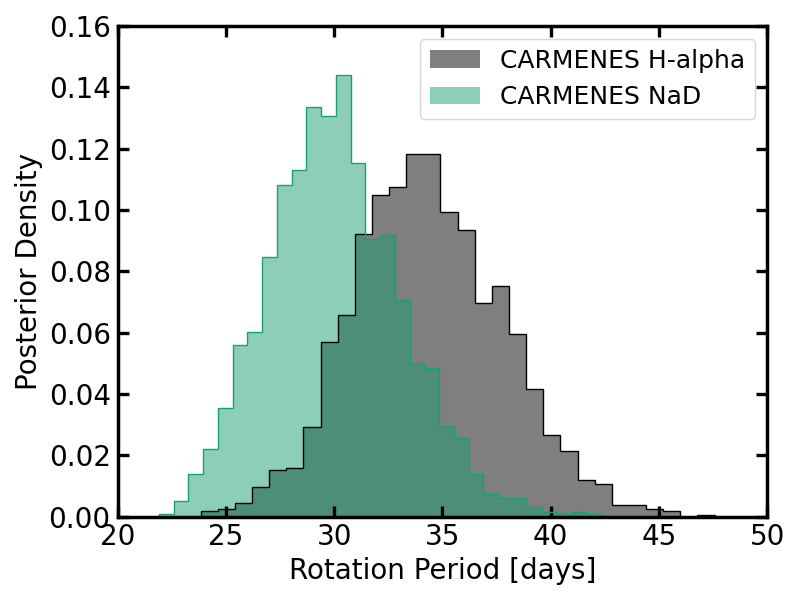}
\caption{The rotation period posterior distribution for K2-18 are depicted. The top panel shows photometric data using datasets from Kepler (purple), ASAS (orange), CATALINA (pink), STELLA, R (green), and the B filter (black). The middle panel shows posterior distribution for activity indicators observed with the HARPS spectrograph, including H-alpha (grey), Na D (green), He I (orange), and Ca I (purple). Finally, the bottom panel shows the posterior distribution for the CARMENES H-alpha (grey) and Na D (green) activity indicator.}
\label{fig:rotationperiod_k218_all}
\end{figure}

\begin{figure}
    \centering
    \includegraphics[width=0.49\textwidth]{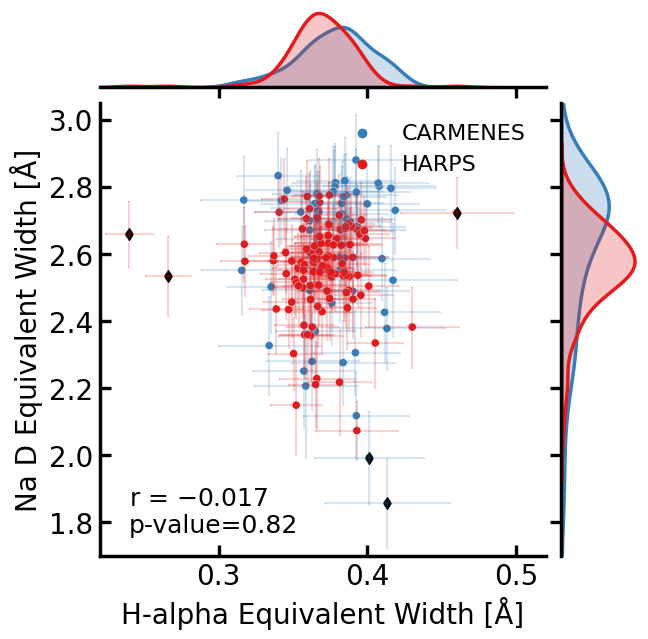}
    \caption{Joint scatter plot of H-alpha  vs. Na D equivalent widths (Å) for K2-18. Data points are coloured by instrument (blue: CARMENES, red: HARPS). Error bars represent measurement uncertainties. Outliers identified by z-scores ($>$3 standard deviations) are plotted as black diamonds. The Pearson coefficient $r$ and the p-value confirm that the correlation is not statistically significant.}
    \label{fig:Nad_halpha}
\end{figure}

\section{Analysis and results} \label{sec:analysis}
We now present the analyses of the data described above for both targets, K2-18 and TOI-732, and the resulting constraints on their properties. We focus on a range of key  stellar properties including the effective temperatures, metallicities, rotation periods, activity, and ages for both targets. 


\subsection{K2-18}
We begin our analysis by focusing on K2-18, an M2.5V dwarf star. As discussed above, the key data sources we use include high-resolution spectroscopy and high-precision timeseries photometry, all largely in the optical. While the spectroscopic data provide constraints on the effective temperature (T$_{\mathrm{eff}}$), metallicity, rotation period and activity indicators, the photometric data provide complementary constraints on the rotation period and activity cycles, along with an age estimate.  

\subsubsection{Stellar properties}

We employed the high-precision method outlined by \cite{Neves_2014} to estimate the stellar temperature and metallicity of K2-18. Their approach utilises pseudo equivalent widths (pEWs) of iron (Fe) lines in the optical wavelength range (530-690 nm) to derive these fundamental stellar properties. \cite{Neves_2014} meticulously select a set of 4104 lines within the specified wavelength range. They measure the pEWs of these lines and correlate them with established calibrations of [Fe/H] and  T$_{\mathrm{eff}}$ \citep{Neves2012, Casagrande2008} derived from stellar atmosphere models \citep{Allard2011, Allard2013}.

Following this method, we analysed the HARPS spectrum of K2-18 and measured the pEWs of the relevant lines. By employing the calibrations provided by \cite{Neves_2014}, we derived a T$_{\mathrm{eff}}$ of 3645$\pm$52 K and a metallicity of [Fe/H] = 0.10$\pm$0.12 dex for K2-18. Our estimated T$_{\mathrm{eff}}$ is marginally higher, but consistent within the 1$\sigma$ uncertainties, compared to recent estimates of T$_{\mathrm{eff}}$ = 3590 $\pm$ 93 K 
from \cite{Hardegree-Ullman2020} and 3547 $\pm$ 85 K from \cite{Hejazi_2024}. 

Additionally, our derived metallicity is in good agreement with previous measurements \citep{Montet_2015, Benneke2017, Sarkis_2018} and consistent with the [Fe/H] = 0.17 $\pm$ 0.10 dex reported by \cite{Hejazi_2024}. While \cite{Hejazi_2024} focused on determining elemental abundances using AutoSpecFit based on IGRINS spectra, our study employed a different approach based on HARPS spectra and pEW measurements to derive fundamental stellar parameters. These wavelength differences, line lists, and analysis techniques can potentially influence the derived stellar parameters.

\subsubsection{Stellar rotation period}\label{sec:rot_k2_18}
We analysed photometric observations of K2-18 obtained through the STELLA (Cousin B and R filters), the Kepler spacecraft, ASAS and CATALINA. First, we use the Lomb-Scargle periodogram, which is effective at characterising periodic signals in unevenly spaced time series data. The periodogram analysis of the photometric data revealed a dominant periodicity of $\sim$39 days with an exceptionally low false alarm probability (FAP) of $10^{-6}$, indicating a robust stellar rotation signal. Figure \ref{fig:rotationperiod_periodogram} shows the periodogram for all the photometric datasets for K2-18. No other significant periodicities emerged after removing this peak, confirming a stable rotation period for K2-18. This stable rotation period is crucial for understanding its atmospheric dynamics and potential for habitability \citep{Gonzalez2014}. 

We employ a Gaussian process \citep[GP;][]{Rasmussen2006} modelling approach to extract rotation periods from the light curves. This method captures the periodic variations in the photometric data obtained from different filters (\citealt{Aigrain2023}. We use periodogram identified periods as priors to help model the underlying light curve more accurately. Next, we construct a probabilistic model using PyMC3 to analyse the timeseries data. This model included parameters for the mean flux, jitter, and a rotation period modeled as a normal distribution. The GP kernel was defined using a combination of the Stochastic Harmonic Oscillator Term (SHOTerm) and a RotationTerm GP kernels to account for periodic behaviour \citep{Foreman2021}. Table \ref{tab:prior} lists the priors used for the GP model. This model is optimised using the Maximum A Posteriori (MAP) solution, ensuring an effective and robust fitting of the GP to the observed data.

We employed Markov Chain Monte Carlo (MCMC) methods to draw samples from the posterior distribution of the model parameters, providing the uncertainties associated with our estimates. The rotation period samples were extracted from the posterior distribution, and the MAP period was calculated along with its associated uncertainties. This process was repeated for multiple datasets to enable a comparative analysis of the estimated rotation periods. The resulting posterior distribution of rotation periods (see Table~\ref{tab:rot_periods}) are illustrated in Fig.~\ref{fig:rotationperiod_k218_all} (top panel), where each dataset is represented by a distinct colour. For K2-18, the posterior distribution of \( P_{\rm rot} \) is more constrained compared to the prior, demonstrating that the data provide an independent constraint on the rotation period.

Additionally, we used Fleck\footnote{\href{https://github.com/bmorris3/fleck}{https://github.com/bmorris3/fleck} } \citep{Morris_2020} to the K2 light curve of K2-18 and employed the Fleck model \citep{Morris_2020} to quantify the spot coverage, i.e., the fraction of the stellar surface covered by spots, as starspots can modulate the stellar brightness during rotation. Our analysis of the Kepler light curve yielded a spot coverage of $f_S = {0.014}^{+0.002}_{-0.001}$. This measured spot coverage appears to be lower than that of most stars reported by \citep{Morris_2020}, suggesting that K2-18 may exhibit lower activity compared to those stars. Note that the sample in \citet{Morris_2020} primarily consists of FGK stars, while K2-18 is an M dwarf. The different magnetic properties and activity levels of M dwarfs compared to FGK stars may impose limitations on this spot coverage comparison. This measured spot coverage is consistent with the predicted spot coverage, $f_S = {0.010}^{+0.009}_{-0.006}$, based on the spot coverage–stellar age relation from \cite{Morris_2020}, and predicted spot coverage falls within the 1$\sigma$ confidence interval of the measured value from the Kepler data. Note that the age range of stars in the \citet{Morris_2020} sample spans from 10 Myr to 4 Gyr, providing a broad evolutionary context. In comparison, K2-18, with an estimated age of 2.9–3.1 Gyr (see \S\ref{sec:age}), falls toward the older end of this range, potentially influencing its activity level and spot coverage.

We further analysed high-resolution spectroscopic data from the HARPS and CARMENES instruments to complement the photometric study. The activity indicators were computed using ACTIN2\footnote{\href{https://github.com/gomesdasilva/ACTIN2}{https://github.com/gomesdasilva/ACTIN2}} \citep{Gomes2018, Gomes2021}. For HARPS observations, we focus on prominent activity indicators, including the H-alpha, Na D, He I, and Ca I lines. For CARMENES observations, we focus only on the H-alpha and Na D lines, which have sufficient signal-to-noise ratios to estimate the indicators. We employed the same GP modelling approach as used for the photometric data.

 The resulting posterior densities of rotation periods, using activity indicators from HARPS and CARMENES are shown in Fig.~\ref{fig:rotationperiod_k218_all}. The rotation periods derived from the MAP solution are summarised in Table~\ref{tab:rot_periods}. The photometric estimates are in close agreement, with values predominantly around 39.2$\pm$0.5 days. The spectroscopic estimates show some variability, particularly from the CARMENES datasets. However, the majority of the HARPS results are also clustered around the 40 day. Since different filters and indicators trace different layers of the stellar atmosphere, the observed variations in rotation periods across these measurements could be due to differential rotation—where different latitudes of the star rotate at varying speeds. Additionally, the migration of starspots and observational uncertainties might also contribute to these discrepancies. 

\begin{figure*}
    \centering
    \includegraphics[width=\textwidth]{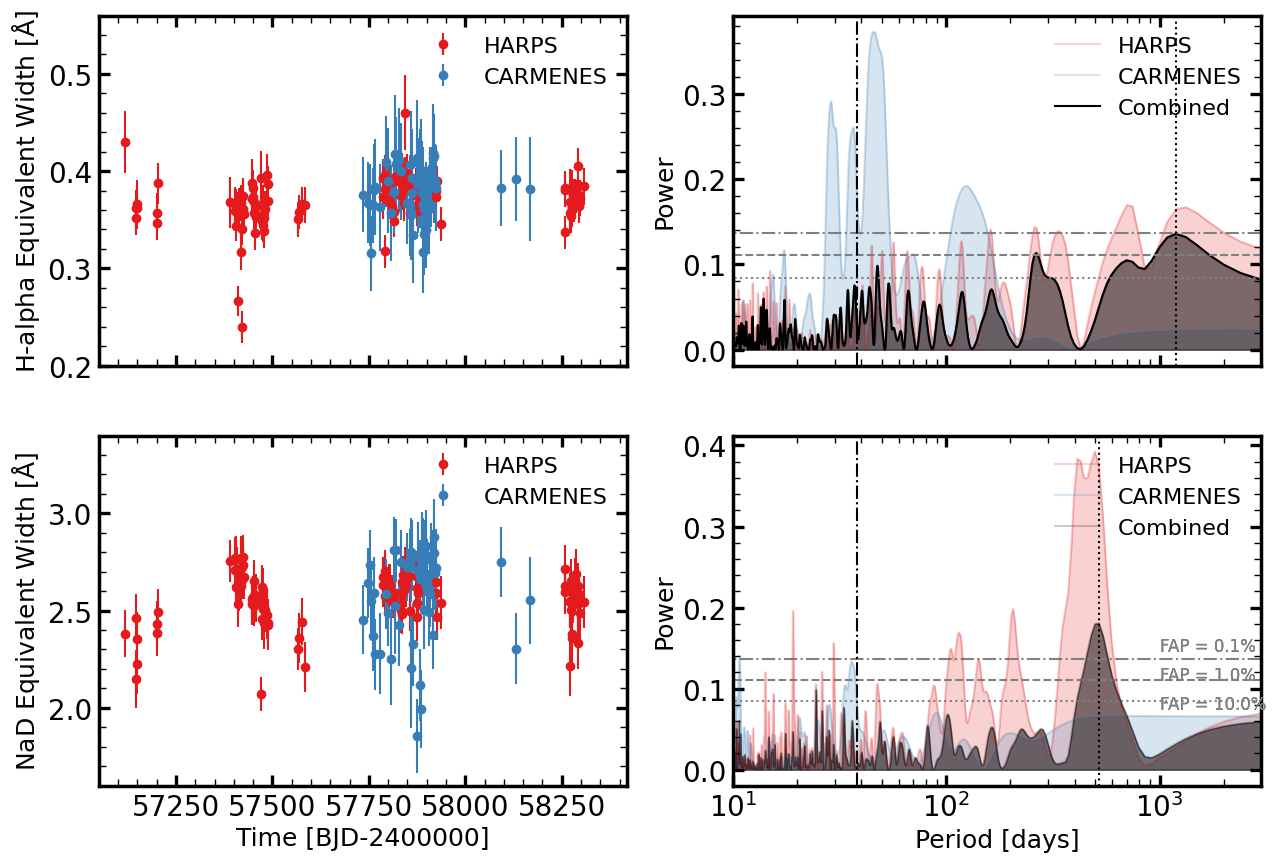}
    \caption{Depicted are the equivalent width time series and Lomb-Scargle periodograms for H-alpha and NaD spectral lines measured with HARPS and CARMENES for K2-18. The left panels show the time series for H-alpha (top) and NaD (bottom) equivalent width. Error bars are included for both instruments (red circles for HARPS and blue circles for CARMENES). The corresponding right panels display the Lomb-Scargle periodograms for H-alpha (top) and NaD (bottom). These plots show the distribution of power across various frequencies. The results for HARPS (red), CARMENES (blue), and the combined data (black) are presented in each panel. The three horizontal dashed lines across the plot correspond to false alarm probabilities of $10\%, 1\%$, and $0.1\%$ The vertical black lines  indicates the stellar rotation period (around 39 days) and the peak period from periodogram analysis. }
    \label{fig:nad_halpha_timeseries}
\end{figure*}

\subsubsection{Stellar activity based on line profile analysis of K2-18}
Spectral lines, particularly those sensitive to chromospheric activity like H-alpha, can vary in strength depending on the level of stellar activity \citep{Robinson1990, Strassmeier1990, Santos2010, Gomes2011, Sissa2016}. Increased activity often leads to stronger chromospheric emission lines and deeper absorption lines \citep{Hall2008}. Measuring the equivalent width (EqW) of these lines provides a quantitative way to track these changes and potentially diagnose periods of high or low activity. 
EqW of H-alpha and Na D lines  and their uncertainties were estimated using a Monte Carlo approach,  using the PHEW python package\footnote{\href{https://github.com/CoolStarsCU/PHEW}{https://github.com/CoolStarsCU/PHEW} }.

Figure \ref{fig:Nad_halpha} shows the joint scatter plot of H-alpha and NaD equivalent width for CARMENES (blue) and HARPS (red) datasets. We employ z-scores (deviations from the mean in units of standard deviation) to identify outliers in the data. Points exceeding a threshold of $\pm$3 standard deviations from the mean value, are considered outliers. These outliers potentially indicate the presence of stellar flares or plages during the observations. Such outliers can introduce spurious signatures in exoplanet atmospheric data. However, careful analysis of outliers, along with an understanding of stellar activity, can provide valuable insights for refining data analysis techniques and achieving more accurate characterisations of exoplanetary atmospheres. 

Figure \ref{fig:Nad_halpha} shows the Pearson correlation coefficient of $-0.017$ between the Na D and H-alpha equivalent widths. The high p-value of 0.82 indicates that this correlation is not statistically significant. This supports the null hypothesis, that there is no correlation between the two variables. Thus, we conclude that there is no strong evidence for a systematic connection between the Na D and H-alpha equivalent widths in this dataset.

\begin{figure*}
    \centering
    \includegraphics[width=\textwidth]{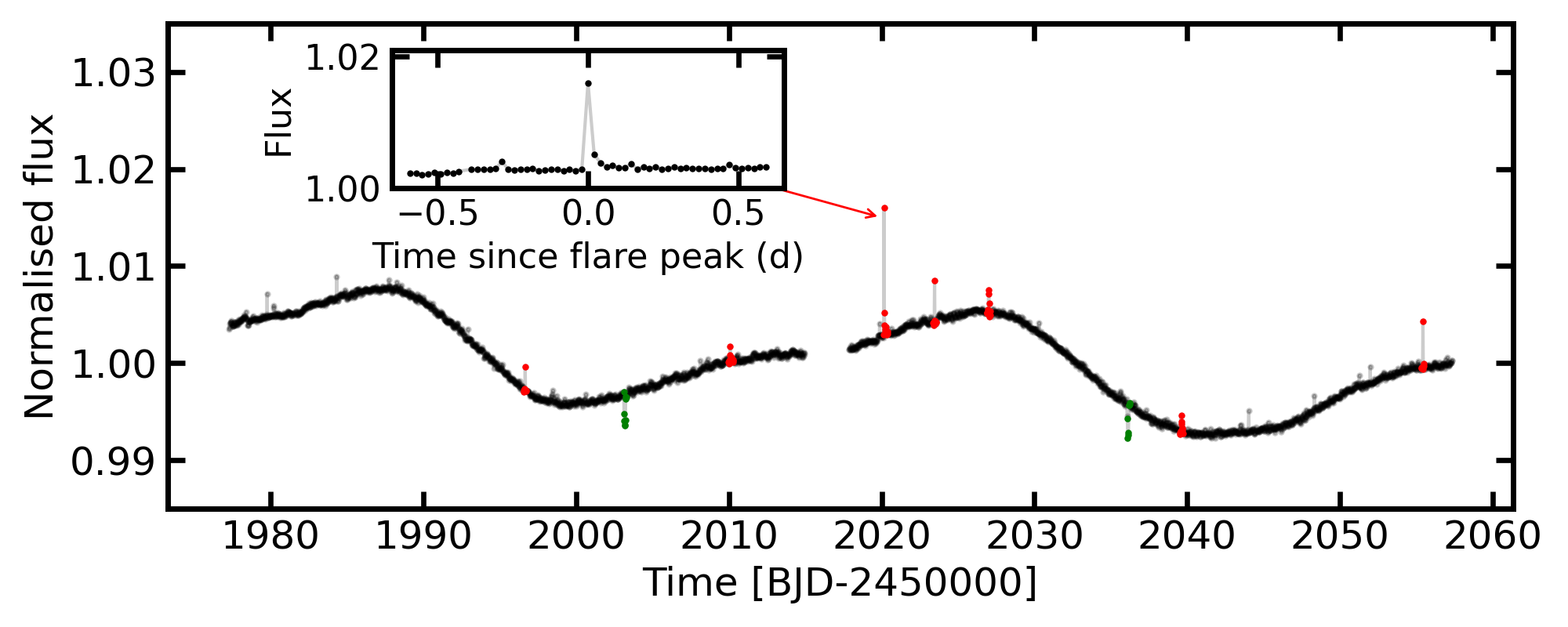}
    \caption{Kepler light curve of K2-18, showing rotational modulation,  flares (marked in red), and planetary transits (marked in green). The inset zooms in on the largest flare, emphasising detailed structures of detected flare events. }
    \label{fig:flare_detection}
\end{figure*}

Figure~\ref{fig:nad_halpha_timeseries} shows the equivalent width time-series for H-alpha and NaD from both HARPS and CARMENES instruments, spanning a combined baseline of approximately 1189 days. While this data is insufficient to conclusively determine the cyclic behaviour of K2-18 using solely spectroscopic data, Lomb-Scargle periodograms were generated for each independent dataset and the combined data to identify potential periodicities. periodogram analysis of the combined dataset revealed a period of 1193$\pm$194 days in H-alpha equivalent width and a period of 518$\pm$25 days in NaD \footnote{Peak period uncertainties were estimated using a parabolic fit to the power peak in the Lomb-Scargle periodogram implemented by  \href{https://pyastronomy.readthedocs.io/en/latest/pyTimingDoc/pyPeriodDoc/gls.html}{PyAstronomy}}. However, visual inspection of the NaD time-series suggests the 518-day period may originate from a specific data segment (BJD 57100-57600). Additionally, closer examination of the NaD lines raises concerns about potential contamination, reducing confidence in this period. Conversely, the period obtained from the H-alpha equivalent width data may be indicative of stellar variability. Future observations of K2-18 in H-alpha using high-precision instruments like HARPS or CARMENES could provide more conclusive evidence.

\subsubsection{Flare activity and high-energy environment of K2-18}

To understand the high-energy environment of K2-18 and its potential impact on the planet K2-18b, we conducted a comprehensive analysis of both flare activity and X-ray emission.
We employed the \texttt{flatwrm} algorithm (\citealt{Vida2018}) to identify flares within the K2 light curve of K2-18b. A total of seven distinct flare events were detected, characterised by a rapid increase in flux followed by a gradual decay. These flares exhibited a wide range of peak fluxes and durations, suggesting varying energy levels. The most prominent flare reached a peak flux of 0.0132, while the weakest flare had a peak flux of 0.0007, measured as the flux difference from the continuum level after light curve normalisation. The full width at half maximum of the flares ranged from 0.0036 to 0.0224 days. Figure \ref{fig:flare_detection} provides a visual representation of the Kepler light curve, highlighting the rotational modulation, flares, and planetary transit. These findings indicate that K2-18b is subjected to a dynamic flaring environment, with the potential for significant energy deposition.

To probe the high-energy environment, we analysed data obtained with the \textit{Chandra X-ray Observatory}'s \citep{Weisskopf2000, Weisskopf2002} Advanced CCD Imaging Spectrometer \citep[ACIS-S;][]{Garmire2003} instrument. The data were processed using standard \textit{Chandra} Interactive Analysis of Observations (CIAO) techniques \citep{Fruscione_2006} with CIAO version 4.16\footnote{\href{https://cxc.harvard.edu/ciao/}{https://cxc.harvard.edu/ciao/}}. The analysis focused on the 0.3-10 keV energy range, excluding higher energies unlikely to be of stellar origin. A circular source region with a radius of $2.5''$ was used for extraction.

The ACIS detector observed K2-18 for a total exposure of 4.03 ks. Due to the low count rate (0.91 counts within the 0.5-7 keV band), a definitive detection of X-ray emission from K2-18 was not possible. Since the source was undetected, we computed the counts required above the background to detect a source with a 2$\sigma$ confidence level. We employed upper limit calculations to constrain the X-ray flux.  
We employed Astrophysical Plasma Emission Code \citep[APEC;][]{Brickhouse2005} with two temperature components (2.5 MK and 5 MK) and solar elemental abundances \citep{Grevesse_1998}. This analysis yielded a model upper-limit X-ray flux of $1.4\times 10^{-14} \text{ erg cm}^{-2} \text{ s}^{-1}$. Considering the distance of $\sim34$ pc, this upper limit on the flux translates to a luminosity upper limit of $2.41 \times 10^{27} \text{ erg s}^{-1}$. 

We checked the consistency of the X-ray luminosity of K2-18 using the X-ray luminosity normalised by stellar surface relation with the age of the star as given by \cite{booth_2017}:

\begin{equation}
 \log_{10}\frac{\mathrm{L_X}}{\mathrm{(R_{\star}/R_{\odot})^2}}=54.65\pm6.98-(2.80\pm0.72) \log_{10}~\mathrm{t}   
 \label{eqn:age_lx}
\end{equation}

where $t$ is the age of the star in Gyr, R$_{\star}$ is the radius of the star in solar radii and L$_{\mathrm{X}}$ is the X-ray luminosity in $\mathrm{erg} \, \mathrm{s}^{-1}$. K2-18, at an age of $2.9 - 3.1 \, \text{Gyr}$ (see \S\ref{sec:age}), with a radius of $0.45 \pm 0.01 \, \mathrm{R_{\odot}}$ \citep{Hardegree-Ullman2020}, produces $\mathrm{L_X} = 2.34 - 2.83 \pm 0.48 \times 10^{27} \, \mathrm{erg} \, \mathrm{s}^{-1}$. The limiting X-ray luminosity measured from \textit{Chandra} is consistent with the luminosity obtained from the age-activity relationship (Eqn. \ref{eqn:age_lx}).

\noindent We calculated the bolometric luminosity (L$_{\mathrm{bol}}$ ) using the formula:
\begin{equation}
  \mathrm{L_{\mathrm{bol}}= 10^{0.4(4.8-\mathrm{m_v-bc_v}+5log(\mathrm{d})-5)} } \mathrm{L}_{\odot}  
\label{eqn:bol}
\end{equation}

\noindent where $\mathrm{m_{v}}$ is the apparent visual magnitude, $\mathrm{bc}_{\mathrm{v}}$ is the bolometric correction for the V-band, $d$ is the distance in parsecs, and L$_{\odot}$ is the solar bolometric luminosity. The term 4.8 in Eqn.~\ref{eqn:bol} serves as the bolometric magnitude of the Sun \citep{cox_2000}. The apparent V magnitude of K2-18 is 13.50$\pm$0.05, and the bolometric correction \( \mathrm{bc_v} = -1.3 \) is determined using the relation from  \cite{Worthey_2011}. This calculation yielded a bolometric luminosity of $1.0 \times 10^{32} \text{ erg s}^{-1}$.

\begin{figure}
    \centering
    \includegraphics[width=0.49\textwidth]{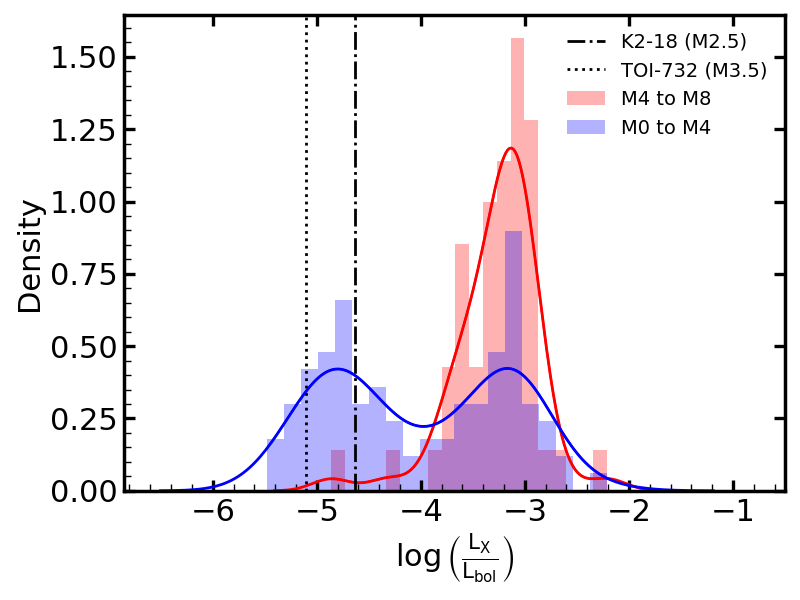}
    \caption{Histogram distribution of 
$\log\mathrm{\frac{L_X}{L_{bol}}}$ for M stars based on data from \citet{Shan2024}. The plot shows histograms and KDE curves for M0–M4 (in blue) and M4–M8 (in red) stars. The vertical lines indicate the upper limits of $\log\mathrm{\frac{L_X}{L_{bol}}}$ for K2-18 (dotted-dashed line) and TOI-732 (dotted line)}.
\label{fig:lxlbol_dist}
\end{figure}

The upper-limit ratio of X-ray to bolometric luminosity ($\log_{10}\mathrm{\frac{L_X}{L_{bol}}}$) was estimated to be $-4.63$. This value places the range typically associated with mild stellar activity for M stars.  This conclusion is based on the $\log_{10}\mathrm{\frac{L_X}{L_{bol}}}$  distribution for M stars, derived from the data in \cite{Shan2024}, as shown in the Figure \ref{fig:lxlbol_dist}. We separate the stars into two spectral type groups: M0 to M4 (in blue) and M4 to M8 (in red). The Kernel Density Estimate (KDE) curves provide a smoothed visualisation of the distribution for M0-M4 and M4-M8  range, overlaid with histograms for comparison. The vertical line depict the upper-limit $\log\mathrm{\frac{L_X}{L_{bol}}}$ for K2-18. Furthermore, as the figure suggests, this upper limit for $\log_{10}\mathrm{\frac{L_X}{L_{bol}}}$ is consistent with the spectral type of K2-18. 
However, given the upper limit, future dedicated X-ray observations may reveal the star to be less active than currently estimated.

Finally, we used the correlation between the rotation period and chromospheric activity level ${\mathrm{\log_{10} R'_{HK}}}$ as given by \cite{Suarez_Mascareno_2018}:
\begin{equation}
 \log_{10}(\mathrm{P}_{\mathrm{rot}})=\mathrm{A+ B}\cdot\log_{10} \mathrm{R}'_{\mathrm{HK}}   
 \label{eqn:prot_rhk}
\end{equation}

where A and B are coefficients given in Table 5 of \cite{Suarez_Mascareno_2018}. Substituting the range of rotation periods of K2-18 in Eqn. \ref{eqn:prot_rhk} gives us a $\mathrm{\log_{10} R'_{HK}}$ between $-5.11$ and $-5.14$. This value of ${\mathrm{\log_{10} R'_{HK}}}$ is further consistent with the star being moderately-low active \citep{Henry1996}.


The upper limit on the X-ray luminosity can be used to place constraints on the extreme ultraviolet (EUV) flux in the $100 - 912\ \text{\AA}$ wavelength range using the scaling relation of \cite{Sanz_2011}. Using Equation 3 from this work, which relates \(\mathrm{L_{\text{EUV}}}\) and L$_{\mathrm{X}}$, we estimate L$_{\mathrm{EUV}} = 2.24 \times 10^{28}\ \mathrm{erg\ s^{-1}}$. This relation also allows extrapolation of the upper limit on the X-ray luminosity to the full XUV range ($5 - 912\ \text{\AA}$). We followed the steps mentioned in \cite{Lalitha2018}, which yielded an upper limit on the XUV luminosity of \(\mathrm{L_{XUV}} \approx 2.48 \times 10^{28}~\text{erg s}^{-1}\). Consequently, the EUV and XUV flux at the position of K2-18b are estimated to be upper limits of $\sim$314 and $\sim$348 erg cm$^{-2}$ s$^{-1}$, respectively. \cite{Santos_2020} measured the EUV flux at the planetary position of K2-18 to be 107.9$^{+124.7}_{-90.8}$ erg cm$^{-2}$ s$^{-1}$. It is important to note that while our estimated upper limit on the EUV flux (approximately 314 erg cm$^{-2}$ s$^{-1}$) is higher than the measured value from \cite{Santos_2020}, it is important to understand the context of this comparison. We can calculate an upper limit based on  \cite{Santos_2020} value, assuming a Gaussian distribution: $107.9 + 3 \times 124.7 = 482$ erg cm$^{-2}$ s$^{-1}$.
While our upper limit of \(\sim314\) erg cm\(^{-2}\) s\(^{-1}\) is lower than this threshold, it emphasises the need for improved observations to constrain X-ray emissions and provide a more robust estimate of the EUV flux.
Future observations with longer exposure times would be beneficial to constrain the X-ray emission and provide a more robust estimate of the EUV flux.

We employ an energy-limited hydrodynamic mass loss model \citep{Watson_1981, Lammer2003, Erkaev_2007, Sanz_2011} to estimate the mass loss rate of K2-18b:
\begin{equation}
   \dot{\mathrm{M}} =\mathrm{\frac{\pi R_p^3 \epsilon F_{XUV}}{G K M_p}}
\label{eqn:mass_loss}
\end{equation}

where $\mathrm{R_p}$ is the planetary radius, $\epsilon=0.4$ is the heating efficiency as suggested by \cite{Valencia2010}, $\mathrm{F_{XUV}}$ is the incident XUV plux at planetary position, G is the gravitational constant, M$_{\mathrm{p}}$ is the mass of the planet, and K is a dimensionless factor which accounts for Roche-lobe filling. We adopt \( K = 1 \), as suggested by \cite{Valencia2010}, which applies to irradiated rocky planets, where atmospheric replenishment from surface sublimation can outpace atmospheric erosion.
Substituting the XUV flux in Eqn: \ref{eqn:mass_loss}, yields a current mass loss rate of approximately $1.9\times10^{8}$ g s$^{-1}$. These steps are similar to those outlined in \cite{Lalitha2014}. We note, that the energy-limited hydrodynamic mass loss approximation is a simplified approach that does not account for several complex factors, such as stellar wind interactions, magnetic fields, or detailed thermal and chemical processes in the planet's atmosphere. As pointed out by \cite{Kubyshkina_2018}, this model provides only a rough estimate of the actual mass loss and should be treated as an upper limit. A more precise calculation would require incorporating these additional physical effects.

\subsection{TOI 732}

We now investigate our second target, TOI-732, an M3.5V star. As discussed in section~\ref{sec:targets_obs}, the key data sources we use for this target include high-resolution spectroscopy in the optical and near-infrared and high-precision time-series photometry in the optical. The stellar properties are derived using the same methods as for K2-18. 

\subsubsection{Stellar properties}

 Following the methodology employed for K2-18, we utilized the high-precision method outlined by \cite{Neves_2014} to estimate the stellar temperature and metallicity of TOI-732. We measured pEWs and subsequently estimated a temperature of 3213$\pm$58 K and a metallicity of [Fe/H] = 0.22$\pm$0.13 dex for TOI-732. In comparison, \cite{Bonfanti2024} used the ODUSSEAS code and HARPS data to derive a temperature of 3358 $\pm$ 92 K and a metallicity of [Fe/H] = 0.06 $\pm$ 0.11 dex. While these values are consistent within the error bars, a potential source of the slight offset could be  due to several factors:
\begin{itemize}
     \item Statistical Treatment: The different statistical methodologies employed in the two analyses may lead to variations in the final parameter estimates. Each method may use distinct approaches to model fitting, error propagation, or uncertainty estimation, which can significantly influence the derived values.
     \item Line Treatment: Although both studies use the same line list from \cite{Neves_2014}, the treatment of specific spectral lines may differ. The difference in how equivalent widths are measured, such as accounting for line blending or saturation effects, could also result in discrepancies in the estimated parameters.
 \end{itemize}

\subsubsection{Stellar rotation period}\label{sec:rot_toi732}

\begin{figure}
    \centering
    \includegraphics[width=0.49\textwidth]{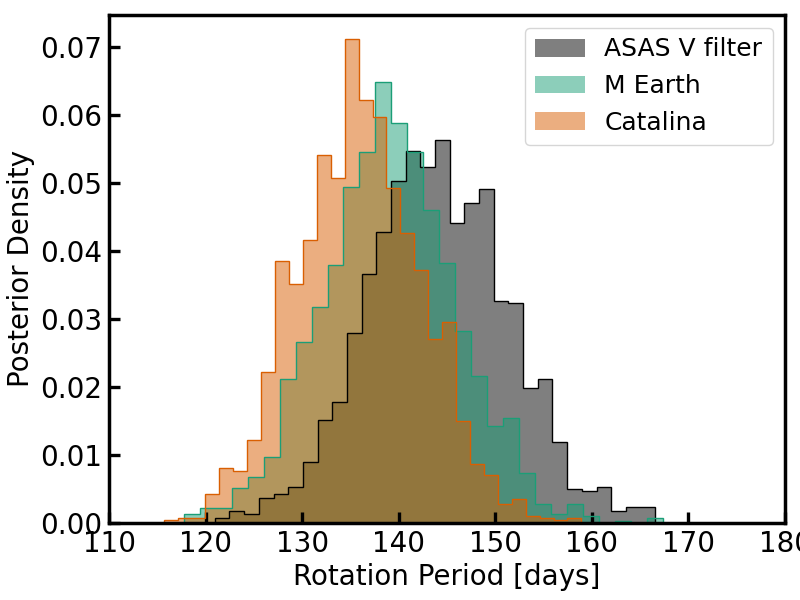}
    \caption{The posterior probability distributions of the rotation period for TOI-732 derived from three independent photometric datasets: ASAS (V filter, orange), CATALINA (blue), and M Earth (green). }
    \label{fig:phot_rotation_toi732}
\end{figure}

We employed the same approach used for K2-18 to determine the rotation period of TOI-732, leveraging data from three extensive photometric surveys: ASAS (V filter), M Earth, and CATALINA, spanning approximately 1508, 1258, and 1889 days, respectively. A GP model, similar to the one used for K2-18, incorporating a rotation term was fitted to the light curves to estimate the posterior distribution of the rotation period. In Table \ref{tab:prior}, we list the priors used for the GP model. Figure~\ref{fig:phot_rotation_toi732} depicts the resulting posterior probability distribution for all three datasets. The analysis revealed peak rotation periods of:

\begin{itemize}
    \item 143 days for the ASAS dataset
    \item 135 days for the CATALINA dataset
    \item 136 days for the M Earth dataset
\end{itemize}

\begin{figure}
    \centering
    \includegraphics[width=0.49\textwidth]{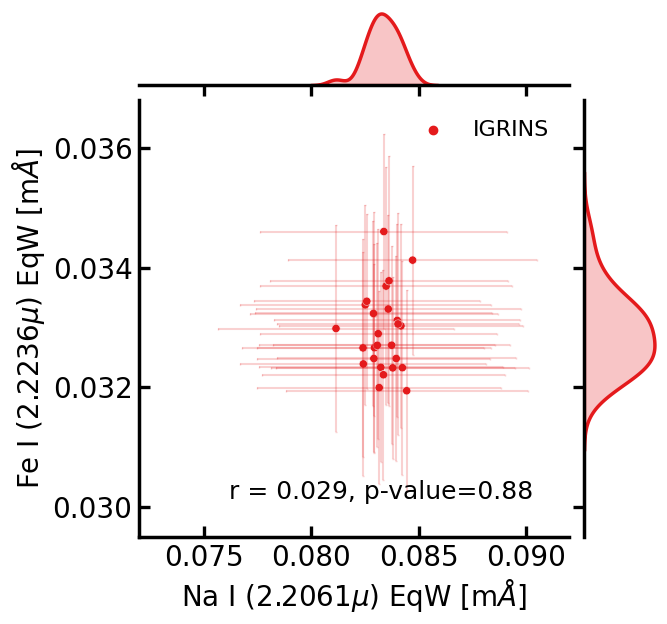}
    \caption{Analysis of the correlation between the Na I line at 2.2061$\mu$m and the Fe I line at 2.2236$\mu$m reveals a weak positive trend for TOI-732. This implies minimal interdependence between the equivalent widths of these two absorption features.}
    \label{fig:igrins_cor_toi732}
\end{figure}

These slight variations in the peak rotation periods may be attributed to factors like the observing wavelength and uneven time coverage. Based on the posterior distributions from all datasets (see Table~\ref{tab:rot_periods}), we conclude that the rotation period of TOI-732 likely lies within a range of 130 to 145 days, which reflects the uncertainty in the measurements.

The HARPS spectroscopic analysis of various activity indicators (H-alpha, Na D, He I, Ca I) does not yield a consistent rotation period (see  Table \ref{tab:rot_periods} and Figure \ref{fig:toi732_correlation} right panel), with derived periods ranging from approximately 58 to 105 days. These discrepancies likely arise from the fact that each spectral line probes different layers of the stellar atmosphere and responds differently to magnetic activity. For instance, H-alpha is more sensitive to chromospheric activity, while Na D and Ca I may respond to photospheric phenomena. Additionally, stellar surface inhomogeneities such as star spots and plages could affect each indicator differently. Additionally, spectral line blending could play a significant role, particularly in the Na D lines, where we observe spikes that suggest potential contamination. Given this inconsistency and the challenges posed by line blending, we refrain from using these spectroscopic indicators to estimate the rotation period of TOI-732 at this stage.

\subsubsection{Stellar activity based on line profile analysis of TOI-732}
Our investigation into chromospheric activity indicators, such as the EqW of H-alpha and Na D lines using the HARPS and CARMENES spectra (see Appendix Figure~\ref{fig:toi732_correlation}). The analysis revealed a weak negative correlation between H-alpha and Na D equivalent width (r = $-0.21$, p = 0.06), suggesting a potential inverse relationship. However, this correlation may be invalid due to strong contamination in the CARMENES data and slight contamination in the HARPS data affecting the estimation of equivalent widths of Na D, unless these lines are appropriately treated. Additionally, the absence of significant long term trends in the spectral lines, observed over a period of 247 d (HARPS coverage), points towards a low activity state for the star (see Appendix Figure~\ref{fig:toi732_ts}).

The relationship between the rotation period and the chromospheric activity level, ${\mathrm{log_{10} R'_{HK}}}$ (\citealt{Suarez_Mascareno_2018}), indicates a value between $-5.49$ and $-5.71$ for TOI-732. This value of ${\mathrm{\log_{10} R'_{HK}}}$ is consistent with the star being inactive \citep{Henry1996}.

In this study, we used IGRINS observations of TOI-732 to investigate a possible relationship between two spectral lines: the Na I line at 2.2061 microns and the Fe I line at 2.2236 microns (see Appendix Figure~\ref{fig:igrins_spec_line}). We chose IGRINS because it covers the spectral range suitable for the comparison of these lines. We aimed to explore whether there is a correlation between Fe I and Na I that is related to stellar activity, similar to the observed relationships for the H-alpha and Na D lines. Figure~\ref{fig:igrins_cor_toi732} describes the analysis of the correlation between the equivalent widths of the Na I and Fe I lines. The calculated Pearson correlation coefficient was 0.029, and the corresponding p-value was 0.88. This result suggests that the correlation between the lines is not statistically significant given the observed data, and we acknowledge that the limited timespan of the data restricts our ability to identify any significant trends or correlations between the strengths of these lines.

\subsubsection{High-energy environment of TOI-732}
We estimate the upper limit of TOI-732's X-ray flux using data from the ACIS instrument on the \textit{Chandra} X-ray Observatory. Since the source was not detected, we calculated the number of counts required above the background to detect a source with a 2$\sigma$ probability. Based on this calculation, we derived an upper limit for the X-ray flux of 9$\times$10$^{-15}$ erg s$^{-1}$ cm$^{-2}$ using an APEC model, similar to the approach used for K2-18. The inferred X-ray luminosity (log L$_{\mathrm{X}}$), given the distance of 22.027 pc, translates to 5.12$\times 10^{26}$ erg s$^{-1}$.

Similar to K2-18, we checked the consistency of the X-ray luminosity using the relationship between the stellar radius, age, and the X-ray luminosity (Eqn. \ref{eqn:age_lx}). Given the radius of $0.380 \pm 0.012 \, \mathrm{R}_{\odot}$ \citep{Bonfanti2024} and an age of $6.9 - 8.1 \, \mathrm{Gyr}$ (see \S\ref{sec:age}), we estimate the X-ray luminosity to be $1.71 - 2.06 \pm 0.17 \times 10^{27} \, \mathrm{erg} \, \mathrm{s}^{-1}$. Although the upper-limit X-ray luminosity is significantly lower than the lower bound of the range derived from stellar parameters, this discrepancy could indicate that the star may be more X-ray bright than the upper-limit counts calculated here. This can only be confirmed by a dedicated X-ray campaign of TOI-732.

The ratio of upper-limit X-ray luminosity to bolometric luminosity $\mathrm{\frac{L_X}{L_{bol}}}$ for TOI-732 is estimated to be around $-5.10$, which falls within the range observed for low-activity M dwarfs (see Fig. \ref{fig:lxlbol_dist}). Furthermore, similar to K2-18, we estimated the upper limit for the XUV flux received by the planet TOI-732c to be $\sim$403 erg s$^{-1}$ cm$^{-2}$. 
 To put this irradiation into context, we compare it with the current XUV irradiation of the Earth, which is \(\text{F}_{\text{XUV, Earth}} = 4.1 \, \mathrm{erg \, s^{-1} \, cm^{-2}}\), and GJ 1214 b, a super-Earth planet orbiting an M4.5 host, which receives approximately \(\text{F}_{\text{XUV, GJ 1214 b}} = 2150 \, \mathrm{erg \, s^{-1} \, cm^{-2}}\) \citep{Lalitha_2014}. The XUV irradiation of TOI-732c is significantly higher than that of Earth but still lower than that of GJ 1214 b, highlighting the energetic environment surrounding these exoplanets. Finally, the mass loss rate of TOI-732c is estimated to be approximately 1.79-1.80$\times10^{8}$ g s$^{-1}$, following a methodology similar to that used for K2-18b. This upper-limit mass-loss rate estimation is based on the upper limit of the X-ray and XUV flux from the host star. Further observations are needed to validate this estimate.

\subsection{Long term trend based on photometric data}

To characterise the long-term variations associated with stellar activity cycles, we employ a GP model, similar to our approach in the  \S\ref{sec:rot_k2_18}. The priors are listed in Table \ref{tab:prior}. We used Lomb-Scargle periodogram to identify potential long-period periods, which served as priors for the PyMC3 model (as described in \S\ref{sec:rot_k2_18}) constructed to fit the GP to the observed data.  

\begin{figure}
\includegraphics[width=0.495\textwidth]{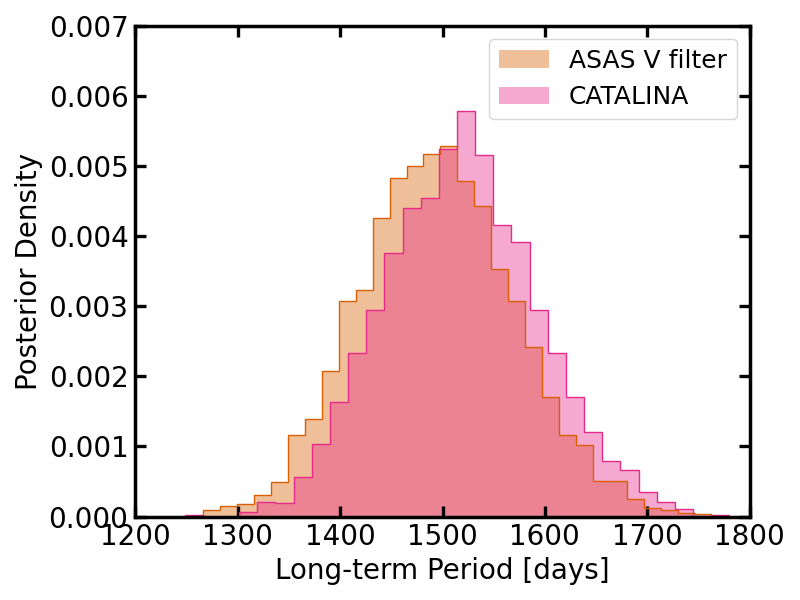}
\caption{Posterior densities of long-term period for ASAS (orange) and CATALINA (pink) datasets for K2-18. This plot illustrates the probability distribution of potential activity cycles in K2-18.}
\label{fig:cycleperiod_photometry}
\end{figure}

For long-term trend analysis of K2-18, we only use ASAS and CATALINA datasets, as we do not use other photometric datasets here, since they lack the long baseline necessary to search for long-term trends. The resulting posterior distributions yield periods of \(1508 \pm 116\) days from ASAS and \(1488 \pm 125\) days from CATALINA. 
These identified periods are shown in Figure \ref{fig:cycleperiod_photometry}, where different colours represent the different datasets.

Based on the potential cyclic and long-period trends identified, we extend this GP framework to model, predict, and forecast the underlying trends in the light curves. We show an example case for K2-18 observed by the ASAS and CATALINA surveys. The GP framework is a non-parametric model, making it well-suited for irregularly sampled data and effectively incorporating uncertainties. We first fit the GP model to the ASAS V-band data to capture the periodic behaviour, estimating parameters such as amplitude, periodicity, length scale, and noise level \citep{Aigrain2023}.

We implemented the PyMC3-based GP model, which used a constant mean and a combination of exponential quadratic covariances with a warped input for periodicity \citep{GONCALVES2020677}. We then optimised the model using MCMC sampling via the PyMC3 framework. After determining the best-fit parameters, we relied on the predictive nature of the GP model to generate forecasts, 
effectively predicting the light curve within a specific time period. Subsequently, we applied the same methodology to the CATALINA data, treating it as a test set. We leveraged the information gained from the ASAS model to guide the predictions for the CATALINA data, allowing us to forecast trends in the test set using the previously learned periodicities.
In Figure~\ref{fig:staccato_k2_18}, we show the observed data, binned data, and model predictions for both datasets. The ASAS light curve was binned into 30-day intervals to enhance the visual representation of 
trends over time. However, GP modelling, was performed on the entire ASAS dataset. This approach allows for the retention of the original temporal resolution and variability in the data.

\begin{figure*}
\includegraphics[width=\textwidth]{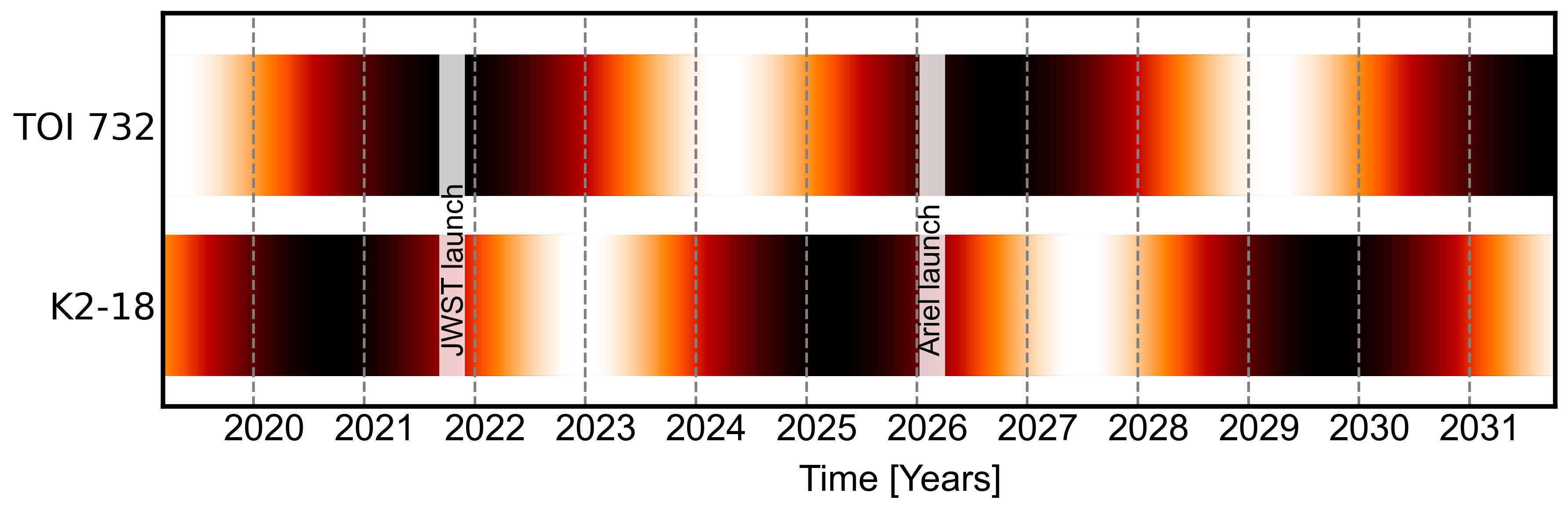}
\caption{Forecast for optimal observations of K2-18 and TOI-732 for the next 10 years. The dark and light regions indicate the duration of activity maxima and minima, respectively. Depicted in grey from left to right are the launches of JWST and the planned launch of ARIEL, respectively. The grey dashed vertical lines represent the start of each year.}
\label{fig:staccato_both}
\end{figure*}

By combining the estimated cycle length with archival photometric data, we can determine the current phase of the activity cycle using the method outlined in \cite{Lalitha_2022}. This approach allows us to forecast the evolution of the activity cycle for our targets over the next 10 years. The results are shown in Figure~\ref{fig:staccato_both}, which highlights the optimal observing windows for our targets during this period. In this figure, the brighter regions indicate times when the star is less active, nearing activity minima, while the darker regions represent periods of greater activity, approaching activity maxima.

According to our forecast, K2-18 was in a low-activity phase during JWST observations in January 2023 and June 2023. While these observations occurred during a predicted low-activity period, residual effects from a recent active phase are still possible. This is supported by the minor spot-crossing event observed in the JWST data, as reported by \cite{Madhusudhan2023}, which aligns with the scenario of K2-18 emerging from a more active phase. The cyclic nature of stellar activity suggests that, K2-18 can be expected to approach a new activity peak in 2025-2026.

For TOI-732, our forecast indicates that the star is currently in a minimal activity phase. As shown in Figure~\ref{fig:staccato_both}, the star is expected to reach its activity maximum between 2026 and 2027. This makes the current period an ideal time for precise atmospheric observations of its exoplanets. Monitoring TOI-732 over the coming years will help us better forecast the optimal observing time for atmospheric characterisation of exoplanets.

To further refine our forecasts and improve the underlying model, it is essential to incorporate new data as it becomes available. By continuously updating the model with fresh observations, we can enhance the accuracy of our predictions and gain a more comprehensive understanding of the star's activity cycle. This iterative process will enable us to refine our forecasts and identify potential discrepancies between our predictions and actual observations, leading to further model improvements.

\begin{figure*}
    \centering
    \includegraphics[width=0.475\textwidth]{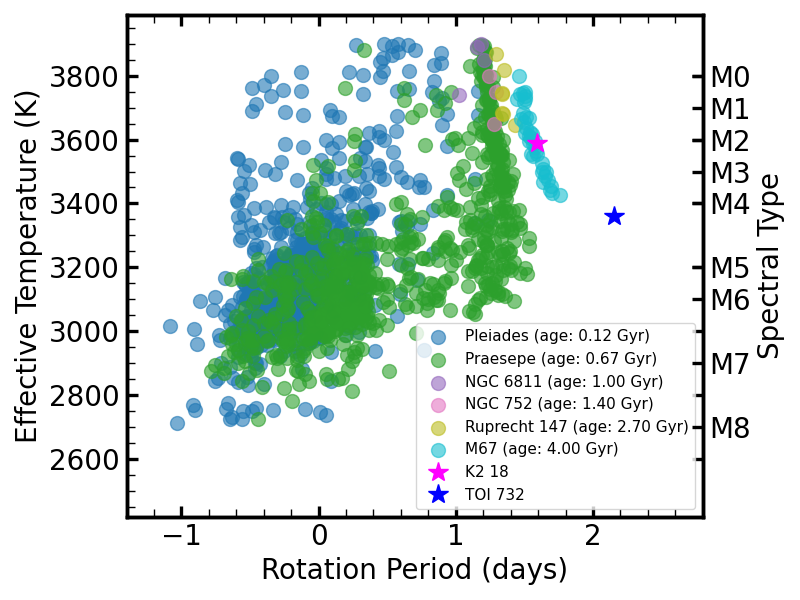}
    \includegraphics[width=0.475\textwidth]{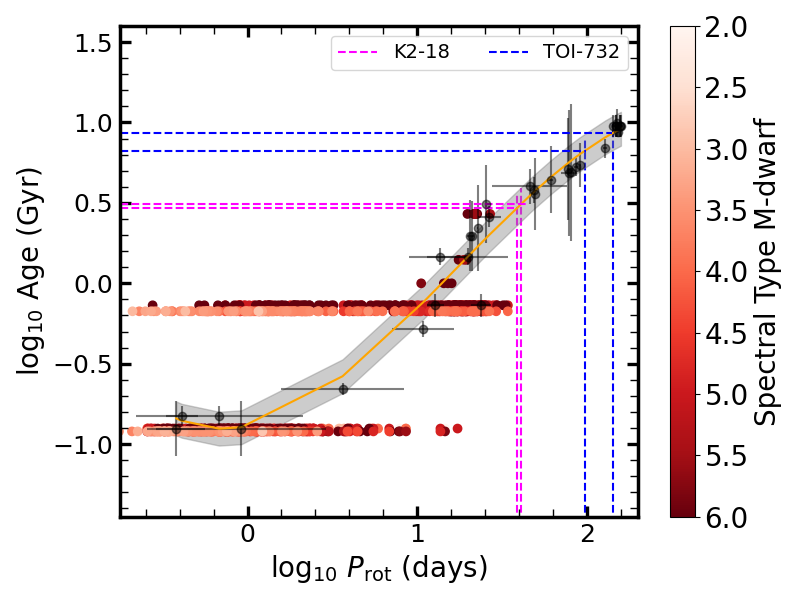}
    \caption{Left panel: The rotation period-effective temperature distribution for M dwarfs across open clusters. Each data point is colour-coded according to its corresponding cluster. The spectral type is indicated on a second y-axis. 
    The filled circles are coloured according to their corresponding cluster membership: 
    Pleiades (age:0.12 Gyr) in blue; Praesepe (age: 0.67 Gyr) in green; NGC 6811 (age: 1 Gyr) in purple; NGC 752 (age: 1.4 Gyr) in pink; Ruprecht 147 (age: 2.7 Gyr) in yellow; M67 (age: 4 Gyr) in cyan. K2-18  and TOI 732  are shown as magenta and blue star symbols, respectively. Right panel: Log-log plot of age and rotation period for M dwarfs in the dataset from \citealt{Engle2023} (shown in black circles). The red colour-coded data points represent cluster members (as reported by \citealt{Curtis_2020} and \citealt{Godoy-Rivera_2021}) spectral types ranging from M2 to M6, as indicated by the colour bar. A third-degree polynomial best fit (Eqn. \ref{eqn:poly}) to the dataset from \citealt{Engle2023} is shown as an orange line. The grey shaded area represents the $\pm 1\sigma$ confidence interval around the fitted model. The magenta and blue vertical lines show the estimated age ranges as given in Table \ref{tab:stellar_parameters_this_work} for K2-18 and TOI-732, respectively.}
    \label{fig:age_rot_relation}
\end{figure*}

\subsection{Age estimation} \label{sec:age}

Determining the ages of M dwarfs, the most common stellar type in the Galaxy \citep{Henry_2004, Bastian2010, Reyle_2021}, presents a significant challenge. Unlike their Sun-like counterparts, which rely on core hydrogen fusion rates for age estimation, M dwarfs have much slower burning processes due to their lower masses. The main sequence lifetime of M dwarfs far exceeds that of Sun-like stars, with these stars staying on the main sequence stage for approximately $10^{11}$ years \citep{Laughlin1997, Baraffe_1998, Dotter2008}. For instance, stars below 0.2 M$_{\odot}$ could potentially reach ages of up to $\sim$1 trillion years \citep{Choi_2016, Engle2023}.

One of the primary challenges in determining the ages of M dwarfs lies in their slow depletion of lithium (Li) within their cores compared to Sun-like stars \citep{Jeffries2005}. This slower burn rate diminishes the effectiveness of lithium abundance as an age indicator for older M dwarfs \citep{Martin2018}.

M dwarfs are known for their high levels of activity, driven by strong magnetic fields and stellar rotation 
\cite{Shulyak2019}.  This activity manifests as chromospheric emission and can mask age-related changes in stellar properties, hindering age estimation based on these properties \citep{Meunier2024}.

Despite these challenges, indirect methods offer possibilities for estimating M dwarf ages. One such method, gyrochronology, has traditionally been calibrated for FGK stars, but recent and novel works have extended these relations to M dwarfs \citep{Dungee2022, Engle2023, Eric2023, Engle2024}. Younger M dwarfs typically exhibit faster rotation, which gradually slows down over time due to magnetic braking mechanisms \citep{Popinchalk2021}. Another approach involves membership in stellar associations or clusters. This method is generally limited to young stars, as the age of an M dwarf can be approximated based on the well-defined age of the cluster \citep{Gruner2020, Curtis_2020}. Studies of nearby young clusters harbouring M dwarfs are currently refining the gyrochronology relationship for these stars, paving the way for more accurate age estimations \citep{Engle2023}.

\cite{Engle2023} investigated the relationship between rotation rate and age for M dwarfs (spectral types M0 to M6.5) as part of their "Living with a Red Dwarf" program. They achieved this by analysing data from stellar clusters, where stars are formed at roughly the same time. 
In Figure~\ref{fig:age_rot_relation} (left panel) we show the effective temperature or the spectral type distribution as a function of rotation period distribution for a given age of the open cluster obtained from Gaia DR2. This plot demonstrates the key concept in gyrochronology: as stars age, they lose angular momentum through mechanisms like magnetic braking. This loss of angular momentum manifests as a gradual decrease in rotation rate over time. Therefore, the left side of the plot, representing younger clusters, exhibits a higher concentration of M dwarfs with faster rotation periods. Conversely, the right side of the plot, populated by older clusters, shows a predominance of M dwarfs with slower rotation rates.

To quantify this relationship, we re-analysed the dependence of stellar rotation on its age using the M dwarfs dataset from \cite{Engle2023}. We used the Bayesian information Criterion \citep[BIC;][]{Schwarz1978} to determine the best-fit polynomial degree for the relationship. The BIC balances the goodness of fit with model complexity, penalising models with more parameters. Lower BIC values indicate a better fit. We found that a third-degree polynomial provided the best fit to the data. The equation for the polynomial fit is given by:

\begin{equation}
\begin{aligned}
\log_{10}(\text{Age}) = \ & a_0 + a_1 (\log_{10}\text{P}_{\text{rot}})  + a_2 (\log_{10}\text{P}_{\text{rot}})^2 \\
& + a_3 (\log_{10}P_{\text{rot}})^3  
\end{aligned}
\label{eqn:poly}
\end{equation}

Here, the age is measured in \textit{Gyr} (gigayears), and the rotation period P$_{\text{rot}}$ is expressed in \textit{days}. The coefficients of the polynomial along with their respective uncertainties are listed in the Table \ref{tab:coefficients}.

\begin{table}
    \centering
        \caption{Best fitting coefficients for the third order polynomial function  Eqn. \ref{eqn:poly} relating $\log_{10}(\text{Age})$ and $\log_{10}(P_{\text{rot}})$, with their uncertainties at 1$\sigma$ level.}
    \begin{tabular}{cccc}
        \hline
        \textbf{Coefficient} & \textbf{Value} & \textbf{Uncertainty} \\
        \hline
        $a_0$ & $0.8880$ & $\pm 0.0613$ \\
        $a_1$ & $0.2176$ & $\pm 0.1260$ \\
        $a_2$ & $0.7087$ & $\pm 0.1915$ \\
        $a_3$ & $-0.1934$ & $\pm 0.0615$ \\
        \hline
    \end{tabular}

    \label{tab:coefficients}
\end{table}

Figure~\ref{fig:age_rot_relation} (right panel) presents the dataset from \citet{Engle2023} (black markers) in the age-P$_{\text{rot}}$ space. We also include M dwarfs from stellar clusters, as reported by \citet{Curtis_2020} and \citet{Godoy-Rivera_2021}, which are color-coded based on the spectral types of the stars. The orange line represents the best-fit third-degree polynomial model described by Eq.~\ref{eqn:poly}. The grey shaded area depicts the estimated $\pm1\sigma$ confidence interval surrounding the fitted model.

We applied the derived third-degree polynomial model to estimate the age of K2-18. However, it is important to consider that K2-18's rotation period varies photometrically between 38.9 and 39.7 days depending on the photometric filters and chromospheric activity indicator used. To account for this variability, we estimated the age of K2-18 using both ends of this rotation period range as input to the model. This resulted in an estimated age range of 2.9 to 3.1 Gyr for K2-18. This is consistent with the estimate of 2.4 $\pm$0.6 Gyr by \citet{Guinan_2019} and slightly lower compared to the 3.3-3.4 Gyr range obtained using equations (3) and (4) from \citet{Engle2023}. This difference likely stems from our use of a single polynomial model encompassing all rotation periods, as opposed to their approach of employing separate linear fits for distinct rotational tracks (M2.5–6.5 dwarfs). While their method might account for track-specific trends, our simpler approach with a single polynomial applied to the full rotation range yields a comparable age range for K2-18.

Similarly, we applied the polynomial model to estimate the age of TOI-732. TOI-732's rotation period varies between 97 and 143 days. Using these rotation period values as input to the model, we estimated the age of TOI-732 to be between 6.7 and 8.6 Gyr. This estimated age range is consistent with the range obtained using the \cite{Engle2023} relation (5.7 to 9.0 Gyr).

\begin{table}
\caption{Summary of stellar parameters for K2-18 and TOI-732 systems derived in this work.}
\label{tab:stellar_parameters_this_work}
\begin{tabular}{llll}
\hline
\textbf{Parameter} & \textbf{K2-18} & \textbf{TOI-732} \\
\hline
T$_{\mathrm {eff}}$ (K) & 3645$\pm$52 & 3213$\pm$92\\
Metallicity [Fe/H]  & 0.10$\pm$0.12&0.22$\pm$0.13\\
P$_{\mathrm{rot}} ~$(d) & 38.9--40.2 & 135--143\\
${\mathrm{\log_{10} R'_{HK}}}$ &  $-5.11$ -- $-5.14$ &  $-5.49$ -- $-5.71$ \\
log L$_{\mathrm{X}}$ (erg s$^{-1}$) & 27.38 & 27.64 \\
log(L$_{\mathrm{X}}$/L$_{\mathrm{bol}}$) & $-4.63$ & $-5.10$\\
Age (Gyr) & 2.9--3.1& 6.7--8.6\\
\hline
\end{tabular}

\textbf{Notes:} The rotation periods are computed values from photometric datasets. The log L$_{\mathrm{X}}$ and log(L$_{\mathrm{X}}$/L$_{\mathrm{bol}}$) are upper limit values.
\end{table}

\section{Summary and Discussion}\label{sec:summary}

Low-mass planets orbiting M dwarfs are important targets in the search for exoplanetary habitability and biosignatures. Detailed understanding of such planets relies on accurate characterisation of their host stars which can affect both spectroscopic observations of the planets as well as their planetary processes, including habitability. In this work, we investigated two such M dwarfs, K2-18 and TOI-732, which are known to host candidate Hycean worlds that are promising targets for atmospheric spectroscopy with JWST \citep{Cloutier_2020, Madhusudhan_2021, Savvas2022, Madhusudhan2023}. We utilised high-resolution optical and/or infrared spectroscopy and photometric time-series observations from various facilities to determine a range of properties for each target, including effective temperature, metallicity, rotation periods, and activity levels using established techniques. We also estimate their ages using gyrochronology, based on empirical relations between stellar rotation and age for M dwarfs. A summary of the derived stellar parameters for K2-18 and TOI-732 are presented in Table \ref{tab:stellar_parameters_this_work}.

Our derived stellar parameters are generally consistent with previous studies. We determined effective temperatures of 3645$\pm$52 K and 3213$\pm$92 K for K2-18 and TOI-732, respectively. Our metallicity estimates are 0.10$\pm$0.12 and 0.22$\pm$0.13 dex for K2-18 and TOI-732, respectively. TOI-732 shows a marginally higher metallicity compared to K2-18, suggesting potentially higher enrichment of heavy elements in its planets.

Our analysis indicated that both K2-18 and TOI-732 are relatively quiescent stars. Characterised by low levels of stellar activity, these systems exhibited minimal chromospheric activity as evidenced by weak correlations between spectral line indicators. The upper-limit X-ray luminosities further support this classification. The rotational periods are found to be about 40 days for K2-18 and 140 days for TOI-732, resulting in gyrochronological age estimates of 2.9-3.1 Gyr and 6.7-8.6 Gyr, respectively. These stellar properties could provide important inputs for modelling the atmospheric evolution of planets in these systems.

While this work utilises both optical and near-infrared data, further exploration using near-infrared ground-based observations may hold significant promise for characterising stellar activity.
Although some studies have explored infrared signatures of stellar activity, our understanding of these relationships remains limited \citep{Wise_2018, Cortes2023}. By examining the near-infrared counterparts of optical activity-sensitive lines, we have the potential to identify new activity indicators and understand the physical processes driving stellar activity. Additionally, exploring new avenues such as Doppler imaging to infer surface maps could provide information on stellar magnetic fields and activity \citep{Vogt1987, Luger_2021}.These techniques, including Zeeman Doppler Imaging, are effective in probing stellar magnetic fields \citep{Semel1989}. Advances in magnetic field reconstruction have been demonstrated in studies such as \citep[e.g.][]{Marsh1988, Hussain2000, Kochukhov2002, Donati2006, Rosen2015}. Such constraints could potentially provide insight into the effects of magnetic fields and stellar winds on the planetary atmospheres \citep[e.g.][]{Evensberget2023, RMM2019}, especially with regard to potential hycean worlds around M dwarfs (\citealt{Madhusudhan_2021}) as considered in the present work.

Accurately forecasting stellar activity cycles is also important for planning exoplanet observations. While our analysis suggested that K2-18 might currently be at a minimum in its cycle, residual effects from a recent period of elevated activity cannot be entirely ruled out. This is evident in the spot crossing event observed in JWST data reported by \cite{Madhusudhan2023}, aligning with the scenario of a star emerging from a more active phase. The possibility of a stellar activity cycle in K2-18 makes it important to consider stellar activity variations during future atmospheric observations of K2-18 b. We predict potential activity peaks for K2-18 around 2025-2026 and for TOI-732 between 2026-2028. However, it is essential to note that these forecasts are based on modelling currently available data. By continuously monitoring stellar activity and refining the forecasting models, we can improve the accuracy of our predictions and ensure that exoplanet observations are planned during periods of minimal stellar activity, thereby potentially enhancing the reliability of atmospheric characterisation. To further mitigate the impact of stellar activity on exoplanet observations, coordinated ground-based and space-based monitoring could be very helpful. Ground-based telescopes can monitor stellar activity levels complementary to space-based observations, providing a more comprehensive picture. This combined approach could help disentangle the effects of stellar activity from the true signatures of exoplanetary atmospheres.

Overall, we find both K2-18 and TOI-732 to be promising targets for atmospheric characterisation of their planets due to their low levels of stellar activity. However, given the expected variation of activity levels over time, long-term monitoring utilising complementary ground-based and space-based facilities would be helpful to account for potential stellar activity variations and achieve a comprehensive understanding of the stars and their planetary systems.

\section*{Acknowledgements}
This work is supported by the UK Research and Innovation (UKRI) Frontier Research Grant EP/X025179/1 (PI: Nikku Madhusudhan).

\section*{Data Availability}

The spectroscopic data analysed in this study are publicly available from the following archives: 
HARPS can be downloaded from the ESO public archive \footnote{\href{http://archive.eso.org/}{http://archive.eso.org/}}; HIRES Data from the Keck Observatory Archive \footnote{\href{https://koa.ipac.caltech.edu/}{https://koa.ipac.caltech.edu/}}; 
CARMENES Data from the CARMENES archive \footnote{\href{http://carmenes.cab.inta-csic.es/gto/jsp/dr1Public.jsp}{http://carmenes.cab.inta-csic.es/gto/jsp/dr1Public.jsp}};  IGRINS from  the Gemini Observatory Archive \footnote{\href{https://archive.gemini.edu/searchform}{https://archive.gemini.edu/searchform}}. The photometric data we used are available for download at the websites of ASAS\footnote{\href{https://www.astrouw.edu.pl/asas/}{https://www.astrouw.edu.pl/asas/}}, CATALINA \footnote{\href{http://nesssi.cacr.caltech.edu/DataRelease/index1.html}{http://nesssi.cacr.caltech.edu/DataRelease/index1.html}}, and the MEarth project\footnote{\href{https://lweb.cfa.harvard.edu/MEarth/Welcome.html}{https://lweb.cfa.harvard.edu/MEarth/Welcome.html}}. Kepler data is available to download from the Mikulski Archive for Space Telescopes\footnote{\href{https://mast.stsci.edu/portal/Mashup/Clients/Mast/Portal.html}{https://mast.stsci.edu/portal/Mashup/Clients/Mast/Portal.html}}. STELLA data are digitised from \citep{Sarkis_2018}.



\bibliographystyle{mnras}
\bibliography{mnras} 

\begin{thebibliography}{}
\makeatletter
\relax
\def\mn@urlcharsother{\let\do\@makeother \do\$\do\&\do\#\do\^\do\_\do\%\do\~}
\def\mn@doi{\begingroup\mn@urlcharsother \@ifnextchar [ {\mn@doi@} {\mn@doi@[]}}
\def\mn@doi@[#1]#2{\def\@tempa{#1}\ifx\@tempa\@empty \href {http://dx.doi.org/#2} {doi:#2}\else \href {http://dx.doi.org/#2} {#1}\fi \endgroup}
\def\mn@eprint#1#2{\mn@eprint@#1:#2::\@nil}
\def\mn@eprint@arXiv#1{\href {http://arxiv.org/abs/#1} {{\tt arXiv:#1}}}
\def\mn@eprint@dblp#1{\href {http://dblp.uni-trier.de/rec/bibtex/#1.xml} {dblp:#1}}
\def\mn@eprint@#1:#2:#3:#4\@nil{\def\@tempa {#1}\def\@tempb {#2}\def\@tempc {#3}\ifx \@tempc \@empty \let \@tempc \@tempb \let \@tempb \@tempa \fi \ifx \@tempb \@empty \def\@tempb {arXiv}\fi \@ifundefined {mn@eprint@\@tempb}{\@tempb:\@tempc}{\expandafter \expandafter \csname mn@eprint@\@tempb\endcsname \expandafter{\@tempc}}}

\bibitem[\protect\citeauthoryear{{Aigrain} \& {Foreman-Mackey}}{{Aigrain} \& {Foreman-Mackey}}{2023}]{Aigrain2023}
{Aigrain} S.,  {Foreman-Mackey} D.,  2023, \mn@doi [\araa] {10.1146/annurev-astro-052920-103508}, \href {https://ui.adsabs.harvard.edu/abs/2023ARA&A..61..329A} {61, 329}

\bibitem[\protect\citeauthoryear{{Allard}, {Homeier}  \& {Freytag}}{{Allard} et~al.}{2011}]{Allard2011}
{Allard} F.,  {Homeier} D.,   {Freytag} B.,  2011, in {Johns-Krull} C.,  {Browning} M.~K.,   {West} A.~A.,  eds,  Astronomical Society of the Pacific Conference Series Vol. 448, 16th Cambridge Workshop on Cool Stars, Stellar Systems, and the Sun. p.~91 (\mn@eprint {arXiv} {1011.5405}), \mn@doi{10.48550/arXiv.1011.5405}

\bibitem[\protect\citeauthoryear{{Allard}, {Homeier}, {Freytag}, {Schaffenberger}  \& {Rajpurohit}}{{Allard} et~al.}{2013}]{Allard2013}
{Allard} F.,  {Homeier} D.,  {Freytag} B.,  {Schaffenberger} W.,   {Rajpurohit} A.~S.,  2013, \mn@doi [Memorie della Societa Astronomica Italiana Supplementi] {10.48550/arXiv.1302.6559}, \href {https://ui.adsabs.harvard.edu/abs/2013MSAIS..24..128A} {24, 128}

\bibitem[\protect\citeauthoryear{{Apai} et~al.,}{{Apai} et~al.}{2018}]{Apai2018}
{Apai} D.,  et~al., 2018, \mn@doi [arXiv e-prints] {10.48550/arXiv.1803.08708}, \href {https://ui.adsabs.harvard.edu/abs/2018arXiv180308708A} {p. arXiv:1803.08708}

\bibitem[\protect\citeauthoryear{{Baraffe}, {Chabrier}, {Allard}  \& {Hauschildt}}{{Baraffe} et~al.}{1998}]{Baraffe_1998}
{Baraffe} I.,  {Chabrier} G.,  {Allard} F.,   {Hauschildt} P.~H.,  1998, \mn@doi [\aap] {10.48550/arXiv.astro-ph/9805009}, \href {https://ui.adsabs.harvard.edu/abs/1998A&A...337..403B} {337, 403}

\bibitem[\protect\citeauthoryear{{Barclay}, {Kostov}, {Col{\'o}n}, {Quintana}, {Schlieder}, {Louie}, {Gilbert}  \& {Mullally}}{{Barclay} et~al.}{2021}]{Barclay2021}
{Barclay} T.,  {Kostov} V.~B.,  {Col{\'o}n} K.~D.,  {Quintana} E.~V.,  {Schlieder} J.~E.,  {Louie} D.~R.,  {Gilbert} E.~A.,   {Mullally} S.~E.,  2021, \mn@doi [\aj] {10.3847/1538-3881/ac2824}, \href {https://ui.adsabs.harvard.edu/abs/2021AJ....162..300B} {162, 300}

\bibitem[\protect\citeauthoryear{{Bastian}, {Covey}  \& {Meyer}}{{Bastian} et~al.}{2010}]{Bastian2010}
{Bastian} N.,  {Covey} K.~R.,   {Meyer} M.~R.,  2010, \mn@doi [\araa] {10.1146/annurev-astro-082708-101642}, \href {https://ui.adsabs.harvard.edu/abs/2010ARA&A..48..339B} {48, 339}

\bibitem[\protect\citeauthoryear{{Benneke} et~al.,}{{Benneke} et~al.}{2017}]{Benneke2017}
{Benneke} B.,  et~al., 2017, \mn@doi [\apj] {10.3847/1538-4357/834/2/187}, \href {https://ui.adsabs.harvard.edu/abs/2017ApJ...834..187B} {834, 187}

\bibitem[\protect\citeauthoryear{{Benneke} et~al.,}{{Benneke} et~al.}{2019}]{Benneke2019}
{Benneke} B.,  et~al., 2019, \mn@doi [\apjl] {10.3847/2041-8213/ab59dc}, \href {https://ui.adsabs.harvard.edu/abs/2019ApJ...887L..14B} {887, L14}

\bibitem[\protect\citeauthoryear{{Benz} et~al.,}{{Benz} et~al.}{2021}]{benz2021}
{Benz} W.,  et~al., 2021, \mn@doi [Experimental Astronomy] {10.1007/s10686-020-09679-4}, \href {https://ui.adsabs.harvard.edu/abs/2021ExA....51..109B} {51, 109}

\bibitem[\protect\citeauthoryear{{Blain}, {Charnay}  \& {B{\'e}zard}}{{Blain} et~al.}{2021}]{Blain2021}
{Blain} D.,  {Charnay} B.,   {B{\'e}zard} B.,  2021, \mn@doi [\aap] {10.1051/0004-6361/202039072}, \href {https://ui.adsabs.harvard.edu/abs/2021A\&A...646A..15B} {646, A15}

\bibitem[\protect\citeauthoryear{{Boldt, S.}, {Oshagh, M.}, {Dreizler, S.}, {Mallonn, M.}, {Santos, N. C.}, {Claret, A.}, {Reiners, A.}  \& {Sedaghati, E.}}{{Boldt, S.} et~al.}{2020}]{Boldt_2020}
{Boldt, S.} {Oshagh, M.} {Dreizler, S.} {Mallonn, M.} {Santos, N. C.} {Claret, A.} {Reiners, A.}  {Sedaghati, E.} 2020, \mn@doi [A\&A] {10.1051/0004-6361/201937419}, 635, A123

\bibitem[\protect\citeauthoryear{{Bonfanti} et~al.,}{{Bonfanti} et~al.}{2024}]{Bonfanti2024}
{Bonfanti} A.,  et~al., 2024, \mn@doi [\aap] {10.1051/0004-6361/202348180}, \href {https://ui.adsabs.harvard.edu/abs/2024A&A...682A..66B} {682, A66}

\bibitem[\protect\citeauthoryear{{Booth}, {Poppenhaeger}, {Watson}, {Silva Aguirre}  \& {Wolk}}{{Booth} et~al.}{2017}]{booth_2017}
{Booth} R.~S.,  {Poppenhaeger} K.,  {Watson} C.~A.,  {Silva Aguirre} V.,   {Wolk} S.~J.,  2017, \mn@doi [\mnras] {10.1093/mnras/stx1630}, \href {https://ui.adsabs.harvard.edu/abs/2017MNRAS.471.1012B} {471, 1012}

\bibitem[\protect\citeauthoryear{{Borucki} et~al.,}{{Borucki} et~al.}{2010}]{Borucki_2010}
{Borucki} W.~J.,  et~al., 2010, \mn@doi [Science] {10.1126/science.1185402}, \href {https://ui.adsabs.harvard.edu/abs/2010Sci...327..977B} {327, 977}

\bibitem[\protect\citeauthoryear{{Brickhouse}, {Desai}, {Hoogerwerf}, {Liedahl}  \& {Smith}}{{Brickhouse} et~al.}{2005}]{Brickhouse2005}
{Brickhouse} N.~S.,  {Desai} P.,  {Hoogerwerf} R.,  {Liedahl} D.~A.,   {Smith} R.~K.,  2005, in {Smith} R.,  ed.,  American Institute of Physics Conference Series Vol. 774, X-ray Diagnostics of Astrophysical Plasmas: Theory, Experiment, and Observation. AIP, pp 405--407, \mn@doi{10.1063/1.1960961}

\bibitem[\protect\citeauthoryear{Cabot, Madhusudhan, Constantinou, Valencia, Vos, Masseron  \& Cheverall}{Cabot et~al.}{2024}]{Cabot_2024}
Cabot S. H.~C.,  Madhusudhan N.,  Constantinou S.,  Valencia D.,  Vos J.~M.,  Masseron T.,   Cheverall C.~J.,  2024, \mn@doi [The Astrophysical Journal Letters] {10.3847/2041-8213/ad3828}, 966, L10

\bibitem[\protect\citeauthoryear{{Casagrande}, {Flynn}  \& {Bessell}}{{Casagrande} et~al.}{2008}]{Casagrande2008}
{Casagrande} L.,  {Flynn} C.,   {Bessell} M.,  2008, \mn@doi [\mnras] {10.1111/j.1365-2966.2008.13573.x}, \href {https://ui.adsabs.harvard.edu/abs/2008MNRAS.389..585C} {389, 585}

\bibitem[\protect\citeauthoryear{{Cheverall} \& {Madhusudhan}}{{Cheverall} \& {Madhusudhan}}{2024}]{Cheverall_2024}
{Cheverall} C.~J.,  {Madhusudhan} N.,  2024, \mn@doi [\aj] {10.3847/1538-3881/ad380c}, \href {https://ui.adsabs.harvard.edu/abs/2024AJ....167..272C} {167, 272}

\bibitem[\protect\citeauthoryear{Choi, Dotter, Conroy, Cantiello, Paxton  \& Johnson}{Choi et~al.}{2016}]{Choi_2016}
Choi J.,  Dotter A.,  Conroy C.,  Cantiello M.,  Paxton B.,   Johnson B.~D.,  2016, \mn@doi [The Astrophysical Journal] {10.3847/0004-637X/823/2/102}, 823, 102

\bibitem[\protect\citeauthoryear{{Christensen} et~al.,}{{Christensen} et~al.}{2012}]{Christensen_2012}
{Christensen} E.,  et~al., 2012, in AAS/Division for Planetary Sciences Meeting Abstracts \#44. p. 210.13

\bibitem[\protect\citeauthoryear{{Cloutier, R.} et~al.,}{{Cloutier, R.} et~al.}{2017}]{Cloutier_2017}
{Cloutier, R.} et~al., 2017, \mn@doi [A\&A] {10.1051/0004-6361/201731558}, 608, A35

\bibitem[\protect\citeauthoryear{{Cloutier} et~al.,}{{Cloutier} et~al.}{2019}]{Cloutier2019}
{Cloutier} R.,  et~al., 2019, \mn@doi [\aap] {10.1051/0004-6361/201833995}, \href {https://ui.adsabs.harvard.edu/abs/2019A&A...621A..49C} {621, A49}

\bibitem[\protect\citeauthoryear{Cloutier et~al.,}{Cloutier et~al.}{2020}]{Cloutier_2020}
Cloutier R.,  et~al., 2020, \mn@doi [The Astronomical Journal] {10.3847/1538-3881/ab91c2}, 160, 3

\bibitem[\protect\citeauthoryear{{Constantinou} \& {Madhusudhan}}{{Constantinou} \& {Madhusudhan}}{2022}]{Savvas2022}
{Constantinou} S.,  {Madhusudhan} N.,  2022, \mn@doi [\mnras] {10.1093/mnras/stac1277}, \href {https://ui.adsabs.harvard.edu/abs/2022MNRAS.514.2073C} {514, 2073}

\bibitem[\protect\citeauthoryear{{Cooke} \& {Madhusudhan}}{{Cooke} \& {Madhusudhan}}{2024}]{Cooke2024}
{Cooke} G.~J.,  {Madhusudhan} N.,  2024, \mn@doi [arXiv e-prints] {10.48550/arXiv.2410.07313}, \href {https://ui.adsabs.harvard.edu/abs/2024arXiv241007313C} {p. arXiv:2410.07313}

\bibitem[\protect\citeauthoryear{{Cort{\'e}s-Zuleta} et~al.,}{{Cort{\'e}s-Zuleta} et~al.}{2023}]{Cortes2023}
{Cort{\'e}s-Zuleta} P.,  et~al., 2023, \mn@doi [\aap] {10.1051/0004-6361/202245131}, \href {https://ui.adsabs.harvard.edu/abs/2023A&A...673A..14C} {673, A14}

\bibitem[\protect\citeauthoryear{Cox}{Cox}{2000}]{cox_2000}
Cox A.~N.,  2000, {Allen's Astrophysical Quantities}.
Springer, New York, NY, \url {https://ui.adsabs.harvard.edu/abs/2000asqu.book.....C}

\bibitem[\protect\citeauthoryear{Curtis et~al.,}{Curtis et~al.}{2020}]{Curtis_2020}
Curtis J.~L.,  et~al., 2020, \mn@doi [The Astrophysical Journal] {10.3847/1538-4357/abbf58}, 904, 140

\bibitem[\protect\citeauthoryear{{Donati} et~al.,}{{Donati} et~al.}{2006}]{Donati2006}
{Donati} J.~F.,  et~al., 2006, \mn@doi [\mnras] {10.1111/j.1365-2966.2006.10558.x}, \href {https://ui.adsabs.harvard.edu/abs/2006MNRAS.370..629D} {370, 629}

\bibitem[\protect\citeauthoryear{{Dotter}, {Chaboyer}, {Jevremovi{\'c}}, {Kostov}, {Baron}  \& {Ferguson}}{{Dotter} et~al.}{2008}]{Dotter2008}
{Dotter} A.,  {Chaboyer} B.,  {Jevremovi{\'c}} D.,  {Kostov} V.,  {Baron} E.,   {Ferguson} J.~W.,  2008, \mn@doi [\apjs] {10.1086/589654}, \href {https://ui.adsabs.harvard.edu/abs/2008ApJS..178...89D} {178, 89}

\bibitem[\protect\citeauthoryear{{Dressing} \& {Charbonneau}}{{Dressing} \& {Charbonneau}}{2013}]{Dressing2013}
{Dressing} C.~D.,  {Charbonneau} D.,  2013, \mn@doi [\apj] {10.1088/0004-637X/767/1/95}, \href {https://ui.adsabs.harvard.edu/abs/2013ApJ...767...95D} {767, 95}

\bibitem[\protect\citeauthoryear{{Dungee}, {van Saders}, {Gaidos}, {Chun}, {Garc{\'\i}a}, {Magnier}, {Mathur}  \& {Santos}}{{Dungee} et~al.}{2022}]{Dungee2022}
{Dungee} R.,  {van Saders} J.,  {Gaidos} E.,  {Chun} M.,  {Garc{\'\i}a} R.~A.,  {Magnier} E.~A.,  {Mathur} S.,   {Santos} {\^A}. R.~G.,  2022, \mn@doi [\apj] {10.3847/1538-4357/ac90be}, \href {https://ui.adsabs.harvard.edu/abs/2022ApJ...938..118D} {938, 118}

\bibitem[\protect\citeauthoryear{{Engle}}{{Engle}}{2024}]{Engle2024}
{Engle} S.~G.,  2024, \mn@doi [\apj] {10.3847/1538-4357/ad0840}, \href {https://ui.adsabs.harvard.edu/abs/2024ApJ...960...62E} {960, 62}

\bibitem[\protect\citeauthoryear{{Engle} \& {Guinan}}{{Engle} \& {Guinan}}{2023}]{Engle2023}
{Engle} S.~G.,  {Guinan} E.~F.,  2023, \mn@doi [\apjl] {10.3847/2041-8213/acf472}, \href {https://ui.adsabs.harvard.edu/abs/2023ApJ...954L..50E} {954, L50}

\bibitem[\protect\citeauthoryear{{Erkaev}, {Kulikov}, {Lammer}, {Selsis}, {Langmayr}, {Jaritz}  \& {Biernat}}{{Erkaev} et~al.}{2007}]{Erkaev_2007}
{Erkaev} N.~V.,  {Kulikov} Y.~N.,  {Lammer} H.,  {Selsis} F.,  {Langmayr} D.,  {Jaritz} G.~F.,   {Biernat} H.~K.,  2007, \mn@doi [\aap] {10.1051/0004-6361:20066929}, \href {https://ui.adsabs.harvard.edu/abs/2007A&A...472..329E} {472, 329}

\bibitem[\protect\citeauthoryear{{Estrela}, {Palit}  \& {Valio}}{{Estrela} et~al.}{2020}]{Estrela2020}
{Estrela} R.,  {Palit} S.,   {Valio} A.,  2020, \mn@doi [Astrobiology] {10.1089/ast.2019.2126}, \href {https://ui.adsabs.harvard.edu/abs/2020AsBio..20.1465E} {20, 1465}

\bibitem[\protect\citeauthoryear{{Evensberget} et~al.,}{{Evensberget} et~al.}{2023}]{Evensberget2023}
{Evensberget} D.,  et~al., 2023, \mn@doi [\mnras] {10.1093/mnras/stad1650}, \href {https://ui.adsabs.harvard.edu/abs/2023MNRAS.524.2042E} {524, 2042}

\bibitem[\protect\citeauthoryear{Foreman-Mackey, Montet, Hogg, Morton, Wang  \& Schölkopf}{Foreman-Mackey et~al.}{2015}]{Foreman-Mackey_2015}
Foreman-Mackey D.,  Montet B.~T.,  Hogg D.~W.,  Morton T.~D.,  Wang D.,   Schölkopf B.,  2015, \mn@doi [The Astrophysical Journal] {10.1088/0004-637X/806/2/215}, 806, 215

\bibitem[\protect\citeauthoryear{{Foreman-Mackey} et~al.,}{{Foreman-Mackey} et~al.}{2021a}]{exoplanet:joss}
{Foreman-Mackey} D.,  et~al., 2021a, arXiv e-prints, \href {https://ui.adsabs.harvard.edu/abs/2021arXiv210501994F} {p. arXiv:2105.01994}

\bibitem[\protect\citeauthoryear{{Foreman-Mackey} et~al.,}{{Foreman-Mackey} et~al.}{2021b}]{Foreman2021}
{Foreman-Mackey} D.,  et~al., 2021b, \mn@doi [The Journal of Open Source Software] {10.21105/joss.03285}, \href {https://ui.adsabs.harvard.edu/abs/2021JOSS....6.3285F} {6, 3285}

\bibitem[\protect\citeauthoryear{{Fruscione} et~al.,}{{Fruscione} et~al.}{2006}]{Fruscione_2006}
{Fruscione} A.,  et~al., 2006, in {Silva} D.~R.,  {Doxsey} R.~E.,  eds,  Society of Photo-Optical Instrumentation Engineers (SPIE) Conference Series Vol. 6270, Observatory Operations: Strategies, Processes, and Systems. p. 62701V, \mn@doi{10.1117/12.671760}

\bibitem[\protect\citeauthoryear{{Fulton} et~al.,}{{Fulton} et~al.}{2017}]{Fulton2017}
{Fulton} B.~J.,  et~al., 2017, \mn@doi [\aj] {10.3847/1538-3881/aa80eb}, \href {https://ui.adsabs.harvard.edu/abs/2017AJ....154..109F} {154, 109}

\bibitem[\protect\citeauthoryear{{Gaia Collaboration}}{{Gaia Collaboration}}{2022}]{Gaia2022}
{Gaia Collaboration} 2022, {VizieR Online Data Catalog: Gaia DR3 Part 1. Main source (Gaia Collaboration, 2022)}, {VizieR On-line Data Catalog: I/355. Originally published in: Astron. Astrophys., in prep. (2022)}, \mn@doi{10.26093/cds/vizier.1355}, \url {https://ui.adsabs.harvard.edu/abs/2022yCat.1355....0G}

\bibitem[\protect\citeauthoryear{{Gaidos}, {Claytor}, {Dungee}, {Ali}  \& {Feiden}}{{Gaidos} et~al.}{2023}]{Eric2023}
{Gaidos} E.,  {Claytor} Z.,  {Dungee} R.,  {Ali} A.,   {Feiden} G.~A.,  2023, \mn@doi [\mnras] {10.1093/mnras/stad343}, \href {https://ui.adsabs.harvard.edu/abs/2023MNRAS.520.5283G} {520, 5283}

\bibitem[\protect\citeauthoryear{{Garmire}, {Bautz}, {Ford}, {Nousek}  \& {Ricker}}{{Garmire} et~al.}{2003}]{Garmire2003}
{Garmire} G.~P.,  {Bautz} M.~W.,  {Ford} P.~G.,  {Nousek} J.~A.,   {Ricker} George~R. J.,  2003, in {Truemper} J.~E.,  {Tananbaum} H.~D.,  eds,  Society of Photo-Optical Instrumentation Engineers (SPIE) Conference Series Vol. 4851, X-Ray and Gamma-Ray Telescopes and Instruments for Astronomy.. pp 28--44, \mn@doi{10.1117/12.461599}

\bibitem[\protect\citeauthoryear{Gilbert-Janizek, Meadows  \& Lustig-Yaeger}{Gilbert-Janizek et~al.}{2024}]{Gilbert-Janizek_2024}
Gilbert-Janizek S.,  Meadows V.~S.,   Lustig-Yaeger J.,  2024, \mn@doi [The Planetary Science Journal] {10.3847/PSJ/ad381e}, 5, 148

\bibitem[\protect\citeauthoryear{Glein}{Glein}{2024}]{glein_geochemical_2024}
Glein C.~R.,  2024, \mn@doi [The Astrophysical Journal Letters] {10.3847/2041-8213/ad3079}, 964, L19

\bibitem[\protect\citeauthoryear{Godoy-Rivera, Pinsonneault  \& Rebull}{Godoy-Rivera et~al.}{2021}]{Godoy-Rivera_2021}
Godoy-Rivera D.,  Pinsonneault M.~H.,   Rebull L.~M.,  2021, \mn@doi [The Astrophysical Journal Supplement Series] {10.3847/1538-4365/ac2058}, 257, 46

\bibitem[\protect\citeauthoryear{{Gomes da Silva}, {Santos}, {Bonfils}, {Delfosse}, {Forveille}  \& {Udry}}{{Gomes da Silva} et~al.}{2011}]{Gomes2011}
{Gomes da Silva} J.,  {Santos} N.~C.,  {Bonfils} X.,  {Delfosse} X.,  {Forveille} T.,   {Udry} S.,  2011, \mn@doi [\aap] {10.1051/0004-6361/201116971}, \href {https://ui.adsabs.harvard.edu/abs/2011A&A...534A..30G} {534, A30}

\bibitem[\protect\citeauthoryear{{Gomes da Silva}, {Figueira}, {Santos}  \& {Faria}}{{Gomes da Silva} et~al.}{2018}]{Gomes2018}
{Gomes da Silva} J.,  {Figueira} P.,  {Santos} N.,   {Faria} J.,  2018, \mn@doi [The Journal of Open Source Software] {10.21105/joss.00667}, \href {https://ui.adsabs.harvard.edu/abs/2018JOSS....3..667G} {3, 667}

\bibitem[\protect\citeauthoryear{{Gomes da Silva} et~al.,}{{Gomes da Silva} et~al.}{2021}]{Gomes2021}
{Gomes da Silva} J.,  et~al., 2021, \mn@doi [\aap] {10.1051/0004-6361/202039765}, \href {https://ui.adsabs.harvard.edu/abs/2021A&A...646A..77G} {646, A77}

\bibitem[\protect\citeauthoryear{{Gonzalez}}{{Gonzalez}}{2014}]{Gonzalez2014}
{Gonzalez} G.,  2014, \mn@doi [Life] {10.3390/life4010035}, \href {https://ui.adsabs.harvard.edu/abs/2014Life....4...35G} {4, 35}

\bibitem[\protect\citeauthoryear{{Grevesse} \& {Sauval}}{{Grevesse} \& {Sauval}}{1998}]{Grevesse_1998}
{Grevesse} N.,  {Sauval} A.~J.,  1998, \mn@doi [\ssr] {10.1023/A:1005161325181}, \href {https://ui.adsabs.harvard.edu/abs/1998SSRv...85..161G} {85, 161}

\bibitem[\protect\citeauthoryear{{Gruner} \& {Barnes}}{{Gruner} \& {Barnes}}{2020}]{Gruner2020}
{Gruner} D.,  {Barnes} S.~A.,  2020, \mn@doi [\aap] {10.1051/0004-6361/202038984}, \href {https://ui.adsabs.harvard.edu/abs/2020A&A...644A..16G} {644, A16}

\bibitem[\protect\citeauthoryear{Guinan \& Engle}{Guinan \& Engle}{2019}]{Guinan_2019}
Guinan E.~F.,  Engle S.~G.,  2019, \mn@doi [Research Notes of the AAS] {10.3847/2515-5172/ab6086}, 3, 189

\bibitem[\protect\citeauthoryear{{Hall}}{{Hall}}{2008}]{Hall2008}
{Hall} J.~C.,  2008, \mn@doi [Living Reviews in Solar Physics] {10.12942/lrsp-2008-2}, \href {https://ui.adsabs.harvard.edu/abs/2008LRSP....5....2H} {5, 2}

\bibitem[\protect\citeauthoryear{{Hardegree-Ullman}, {Zink}, {Christiansen}, {Dressing}, {Ciardi}  \& {Schlieder}}{{Hardegree-Ullman} et~al.}{2020}]{Hardegree-Ullman2020}
{Hardegree-Ullman} K.~K.,  {Zink} J.~K.,  {Christiansen} J.~L.,  {Dressing} C.~D.,  {Ciardi} D.~R.,   {Schlieder} J.~E.,  2020, \mn@doi [\apjs] {10.3847/1538-4365/ab7230}, \href {https://ui.adsabs.harvard.edu/abs/2020ApJS..247...28H} {247, 28}

\bibitem[\protect\citeauthoryear{{Hejazi} et~al.,}{{Hejazi} et~al.}{2024}]{Hejazi_2024}
{Hejazi} N.,  et~al., 2024, \mn@doi [arXiv e-prints] {10.48550/arXiv.2407.07869}, \href {https://ui.adsabs.harvard.edu/abs/2024arXiv240707869H} {p. arXiv:2407.07869}

\bibitem[\protect\citeauthoryear{{Henry}, {Soderblom}, {Donahue}  \& {Baliunas}}{{Henry} et~al.}{1996}]{Henry1996}
{Henry} T.~J.,  {Soderblom} D.~R.,  {Donahue} R.~A.,   {Baliunas} S.~L.,  1996, \mn@doi [\aj] {10.1086/117796}, \href {https://ui.adsabs.harvard.edu/abs/1996AJ....111..439H} {111, 439}

\bibitem[\protect\citeauthoryear{Henry, Subasavage, Brown, Beaulieu, Jao  \& Hambly}{Henry et~al.}{2004}]{Henry_2004}
Henry T.~J.,  Subasavage J.~P.,  Brown M.~A.,  Beaulieu T.~D.,  Jao W.-C.,   Hambly N.~C.,  2004, \mn@doi [The Astronomical Journal] {10.1086/425052}, 128, 2460–2473

\bibitem[\protect\citeauthoryear{{Hu}, {Damiano}, {Scheucher}, {Kite}, {Seager}  \& {Rauer}}{{Hu} et~al.}{2021}]{Hu2021}
{Hu} R.,  {Damiano} M.,  {Scheucher} M.,  {Kite} E.,  {Seager} S.,   {Rauer} H.,  2021, \mn@doi [\apjl] {10.3847/2041-8213/ac1f92}, \href {https://ui.adsabs.harvard.edu/abs/2021ApJ...921L...8H} {921, L8}

\bibitem[\protect\citeauthoryear{{Hussain}, {Donati}, {Collier Cameron}  \& {Barnes}}{{Hussain} et~al.}{2000}]{Hussain2000}
{Hussain} G.~A.~J.,  {Donati} J.~F.,  {Collier Cameron} A.,   {Barnes} J.~R.,  2000, \mn@doi [\mnras] {10.1046/j.1365-8711.2000.03573.x}, \href {https://ui.adsabs.harvard.edu/abs/2000MNRAS.318..961H} {318, 961}

\bibitem[\protect\citeauthoryear{Jeffries \& Oliveira}{Jeffries \& Oliveira}{2005}]{Jeffries2005}
Jeffries R.~D.,  Oliveira J.~M.,  2005, \mn@doi [Monthly Notices of the Royal Astronomical Society] {10.1111/j.1365-2966.2005.08820.x}, 358, 13

\bibitem[\protect\citeauthoryear{{Jeong} et~al.,}{{Jeong} et~al.}{2014}]{Jeong_2014}
{Jeong} U.,  et~al., 2014, in {Holland} A.~D.,  {Beletic} J.,  eds,  Society of Photo-Optical Instrumentation Engineers (SPIE) Conference Series Vol. 9154, High Energy, Optical, and Infrared Detectors for Astronomy VI. p. 91541X, \mn@doi{10.1117/12.2055589}

\bibitem[\protect\citeauthoryear{{Kochukhov} \& {Piskunov}}{{Kochukhov} \& {Piskunov}}{2002}]{Kochukhov2002}
{Kochukhov} O.,  {Piskunov} N.,  2002, \mn@doi [\aap] {10.1051/0004-6361:20020300}, \href {https://ui.adsabs.harvard.edu/abs/2002A&A...388..868K} {388, 868}

\bibitem[\protect\citeauthoryear{Kopparapu et~al.,}{Kopparapu et~al.}{2013}]{Kopparapu_2013}
Kopparapu R.~K.,  et~al., 2013, \mn@doi [The Astrophysical Journal] {10.1088/0004-637X/765/2/131}, 765, 131

\bibitem[\protect\citeauthoryear{{Kubyshkina} et~al.,}{{Kubyshkina} et~al.}{2018}]{Kubyshkina_2018}
{Kubyshkina} D.,  et~al., 2018, \mn@doi [\apjl] {10.3847/2041-8213/aae586}, \href {https://ui.adsabs.harvard.edu/abs/2018ApJ...866L..18K} {866, L18}

\bibitem[\protect\citeauthoryear{{Lalitha}, {Poppenhaeger}, {Singh}, {Czesla}  \& {Schmitt}}{{Lalitha} et~al.}{2014a}]{Lalitha2014}
{Lalitha} S.,  {Poppenhaeger} K.,  {Singh} K.~P.,  {Czesla} S.,   {Schmitt} J.~H.~M.~M.,  2014a, \mn@doi [\apjl] {10.1088/2041-8205/790/1/L11}, \href {https://ui.adsabs.harvard.edu/abs/2014ApJ...790L..11L} {790, L11}

\bibitem[\protect\citeauthoryear{{Lalitha}, {Poppenhaeger}, {Singh}, {Czesla}  \& {Schmitt}}{{Lalitha} et~al.}{2014b}]{Lalitha_2014}
{Lalitha} S.,  {Poppenhaeger} K.,  {Singh} K.~P.,  {Czesla} S.,   {Schmitt} J.~H.~M.~M.,  2014b, \mn@doi [\apjl] {10.1088/2041-8205/790/1/L11}, \href {https://ui.adsabs.harvard.edu/abs/2014ApJ...790L..11L} {790, L11}

\bibitem[\protect\citeauthoryear{Lalitha, Schmitt  \& Dash}{Lalitha et~al.}{2018}]{Lalitha2018}
Lalitha S.,  Schmitt J. H. M.~M.,   Dash S.,  2018, \mn@doi [Monthly Notices of the Royal Astronomical Society] {10.1093/mnras/sty732}, 477, 808

\bibitem[\protect\citeauthoryear{{Lammer}, {Selsis}, {Ribas}, {Guinan}, {Bauer}  \& {Weiss}}{{Lammer} et~al.}{2003}]{Lammer2003}
{Lammer} H.,  {Selsis} F.,  {Ribas} I.,  {Guinan} E.~F.,  {Bauer} S.~J.,   {Weiss} W.~W.,  2003, \mn@doi [\apjl] {10.1086/380815}, \href {https://ui.adsabs.harvard.edu/abs/2003ApJ...598L.121L} {598, L121}

\bibitem[\protect\citeauthoryear{{Laughlin}, {Bodenheimer}  \& {Adams}}{{Laughlin} et~al.}{1997}]{Laughlin1997}
{Laughlin} G.,  {Bodenheimer} P.,   {Adams} F.~C.,  1997, \mn@doi [\apj] {10.1086/304125}, \href {https://ui.adsabs.harvard.edu/abs/1997ApJ...482..420L} {482, 420}

\bibitem[\protect\citeauthoryear{{Lee} \& {Gullikson}}{{Lee} \& {Gullikson}}{2017}]{Lee2017}
{Lee} J.-J.,  {Gullikson} K.,  2017, {igrins/plp v2.2.0-alpha.4}, \mn@doi{10.5281/zenodo.438353}

\bibitem[\protect\citeauthoryear{{Lobo}, {Shields}, {Palubski}  \& {Wolf}}{{Lobo} et~al.}{2023}]{Lobo_2023}
{Lobo} A.~H.,  {Shields} A.~L.,  {Palubski} I.~Z.,   {Wolf} E.,  2023, \mn@doi [\apj] {10.3847/1538-4357/aca970}, \href {https://ui.adsabs.harvard.edu/abs/2023ApJ...945..161L} {945, 161}

\bibitem[\protect\citeauthoryear{{Lovis} et~al.,}{{Lovis} et~al.}{2017}]{Lovis2017}
{Lovis} C.,  et~al., 2017, \mn@doi [\aap] {10.1051/0004-6361/201629682}, \href {https://ui.adsabs.harvard.edu/abs/2017A&A...599A..16L} {599, A16}

\bibitem[\protect\citeauthoryear{Lueftinger, Güdel, Saikia, Johnstone, Kulterer, Kochukhov  \& Kislyakova}{Lueftinger et~al.}{2020}]{Lueftinger2020}
Lueftinger T.,  Güdel M.,  Saikia S.~B.,  Johnstone C.,  Kulterer B.,  Kochukhov O.,   Kislyakova K.,  2020, in Elmegreen B.~G.,  Tóth L.~V.,   Güdel M.,  eds,  IAU Symposium Vol. 345, Origins: From the Protosun to the First Steps of Life. pp 181--184, \mn@doi{10.1017/S174392131900293X}

\bibitem[\protect\citeauthoryear{{Luger}, {Bedell}, {Foreman-Mackey}, {Crossfield}, {Zhao}  \& {Hogg}}{{Luger} et~al.}{2021}]{Luger_2021}
{Luger} R.,  {Bedell} M.,  {Foreman-Mackey} D.,  {Crossfield} I. J.~M.,  {Zhao} L.~L.,   {Hogg} D.~W.,  2021, \mn@doi [arXiv e-prints] {10.48550/arXiv.2110.06271}, \href {https://ui.adsabs.harvard.edu/abs/2021arXiv211006271L} {p. arXiv:2110.06271}

\bibitem[\protect\citeauthoryear{{Madhusudhan}, {Nixon}, {Welbanks}, {Piette}  \& {Booth}}{{Madhusudhan} et~al.}{2020}]{Madhusudhan2020}
{Madhusudhan} N.,  {Nixon} M.~C.,  {Welbanks} L.,  {Piette} A. A.~A.,   {Booth} R.~A.,  2020, \mn@doi [\apjl] {10.3847/2041-8213/ab7229}, \href {https://ui.adsabs.harvard.edu/abs/2020ApJ...891L...7M} {891, L7}

\bibitem[\protect\citeauthoryear{Madhusudhan, Piette  \& Constantinou}{Madhusudhan et~al.}{2021}]{Madhusudhan_2021}
Madhusudhan N.,  Piette A. A.~A.,   Constantinou S.,  2021, \mn@doi [The Astrophysical Journal] {10.3847/1538-4357/abfd9c}, 918, 1

\bibitem[\protect\citeauthoryear{{Madhusudhan}, {Moses}, {Rigby}  \& {Barrier}}{{Madhusudhan} et~al.}{2023a}]{Madhusudhan_2023a}
{Madhusudhan} N.,  {Moses} J.~I.,  {Rigby} F.,   {Barrier} E.,  2023a, \mn@doi [Faraday Discussions] {10.1039/D3FD00075C}, \href {https://ui.adsabs.harvard.edu/abs/2023FaDi..245...80M} {245, 80}

\bibitem[\protect\citeauthoryear{{Madhusudhan}, {Sarkar}, {Constantinou}, {Holmberg}, {Piette}  \& {Moses}}{{Madhusudhan} et~al.}{2023b}]{Madhusudhan2023}
{Madhusudhan} N.,  {Sarkar} S.,  {Constantinou} S.,  {Holmberg} M.,  {Piette} A. A.~A.,   {Moses} J.~I.,  2023b, \mn@doi [\apjl] {10.3847/2041-8213/acf577}, \href {https://ui.adsabs.harvard.edu/abs/2023ApJ...956L..13M} {956, L13}

\bibitem[\protect\citeauthoryear{{Marsh} \& {Horne}}{{Marsh} \& {Horne}}{1988}]{Marsh1988}
{Marsh} T.~R.,  {Horne} K.,  1988, \mn@doi [\mnras] {10.1093/mnras/235.1.269}, \href {https://ui.adsabs.harvard.edu/abs/1988MNRAS.235..269M} {235, 269}

\bibitem[\protect\citeauthoryear{{Mart{\'\i}n}, {Lodieu}, {Pavlenko}  \& {B{\'e}jar}}{{Mart{\'\i}n} et~al.}{2018}]{Martin2018}
{Mart{\'\i}n} E.~L.,  {Lodieu} N.,  {Pavlenko} Y.,   {B{\'e}jar} V. J.~S.,  2018, \mn@doi [\apj] {10.3847/1538-4357/aaaeb8}, \href {https://ui.adsabs.harvard.edu/abs/2018ApJ...856...40M} {856, 40}

\bibitem[\protect\citeauthoryear{{Mayor} et~al.,}{{Mayor} et~al.}{2003}]{Mayor2003}
{Mayor} M.,  et~al., 2003, The Messenger, \href {https://ui.adsabs.harvard.edu/abs/2003Msngr.114...20M} {114, 20}

\bibitem[\protect\citeauthoryear{{Meunier}, {Mignon}, {Kretzschmar}  \& {Delfosse}}{{Meunier} et~al.}{2024}]{Meunier2024}
{Meunier} N.,  {Mignon} L.,  {Kretzschmar} M.,   {Delfosse} X.,  2024, \mn@doi [\aap] {10.1051/0004-6361/202347362}, \href {https://ui.adsabs.harvard.edu/abs/2024A&A...684A.106M} {684, A106}

\bibitem[\protect\citeauthoryear{{Montet} et~al.,}{{Montet} et~al.}{2015}]{Montet_2015}
{Montet} B.~T.,  et~al., 2015, \mn@doi [\apj] {10.1088/0004-637X/809/1/25}, \href {https://ui.adsabs.harvard.edu/abs/2015ApJ...809...25M} {809, 25}

\bibitem[\protect\citeauthoryear{Morris}{Morris}{2020}]{Morris_2020}
Morris B.~M.,  2020, \mn@doi [The Astrophysical Journal] {10.3847/1538-4357/ab79a0}, 893, 67

\bibitem[\protect\citeauthoryear{{Neves} et~al.,}{{Neves} et~al.}{2012}]{Neves2012}
{Neves} V.,  et~al., 2012, \mn@doi [\aap] {10.1051/0004-6361/201118115}, \href {https://ui.adsabs.harvard.edu/abs/2012A&A...538A..25N} {538, A25}

\bibitem[\protect\citeauthoryear{{Neves}, {Bonfils}, {Santos}, {Delfosse}, {Forveille}, {Allard}  \& {Udry}}{{Neves} et~al.}{2014}]{Neves_2014}
{Neves} V.,  {Bonfils} X.,  {Santos} N.~C.,  {Delfosse} X.,  {Forveille} T.,  {Allard} F.,   {Udry} S.,  2014, \mn@doi [\aap] {10.1051/0004-6361/201424139}, \href {https://ui.adsabs.harvard.edu/abs/2014A&A...568A.121N} {568, A121}

\bibitem[\protect\citeauthoryear{{Nowak} et~al.,}{{Nowak} et~al.}{2020}]{Nowak2020}
{Nowak} G.,  et~al., 2020, \mn@doi [\aap] {10.1051/0004-6361/202037867}, \href {https://ui.adsabs.harvard.edu/abs/2020A&A...642A.173N} {642, A173}

\bibitem[\protect\citeauthoryear{{Nutzman} \& {Charbonneau}}{{Nutzman} \& {Charbonneau}}{2008}]{Nutzman_2008}
{Nutzman} P.,  {Charbonneau} D.,  2008, \mn@doi [\pasp] {10.1086/533420}, \href {https://ui.adsabs.harvard.edu/abs/2008PASP..120..317N} {120, 317}

\bibitem[\protect\citeauthoryear{{Oshagh, M.}, {Santos, N. C.}, {Ehrenreich, D.}, {Haghighipour, N.}, {Figueira, P.}, {Santerne, A.}  \& {Montalto, M.}}{{Oshagh, M.} et~al.}{2014}]{Oshagh_2014}
{Oshagh, M.} {Santos, N. C.} {Ehrenreich, D.} {Haghighipour, N.} {Figueira, P.} {Santerne, A.}  {Montalto, M.} 2014, \mn@doi [A\&A] {10.1051/0004-6361/201424059}, 568, A99

\bibitem[\protect\citeauthoryear{Owen \& Mohanty}{Owen \& Mohanty}{2016}]{Owen2016}
Owen J.~E.,  Mohanty S.,  2016, \mn@doi [Monthly Notices of the Royal Astronomical Society] {10.1093/mnras/stw959}, 459, 4088

\bibitem[\protect\citeauthoryear{{Park} et~al.,}{{Park} et~al.}{2014}]{Park2014}
{Park} C.,  et~al., 2014, in {Ramsay} S.~K.,  {McLean} I.~S.,   {Takami} H.,  eds,  Society of Photo-Optical Instrumentation Engineers (SPIE) Conference Series Vol. 9147, Ground-based and Airborne Instrumentation for Astronomy V. p. 91471D, \mn@doi{10.1117/12.2056431}

\bibitem[\protect\citeauthoryear{Pinhas, Rackham, Madhusudhan  \& Apai}{Pinhas et~al.}{2018}]{Pinhas2018}
Pinhas A.,  Rackham B.~V.,  Madhusudhan N.,   Apai D.,  2018, \mn@doi [Monthly Notices of the Royal Astronomical Society] {10.1093/mnras/sty2209}, 480, 5314

\bibitem[\protect\citeauthoryear{{Pojmanski}}{{Pojmanski}}{2000}]{Pojmanski_2000}
{Pojmanski} G.,  2000, \mn@doi [\actaa] {10.48550/arXiv.astro-ph/0005236}, \href {https://ui.adsabs.harvard.edu/abs/2000AcA....50..177P} {50, 177}

\bibitem[\protect\citeauthoryear{{Popinchalk}, {Faherty}, {Kiman}, {Gagn{\'e}}, {Curtis}, {Angus}, {Cruz}  \& {Rice}}{{Popinchalk} et~al.}{2021}]{Popinchalk2021}
{Popinchalk} M.,  {Faherty} J.~K.,  {Kiman} R.,  {Gagn{\'e}} J.,  {Curtis} J.~L.,  {Angus} R.,  {Cruz} K.~L.,   {Rice} E.~L.,  2021, \mn@doi [\apj] {10.3847/1538-4357/ac0444}, \href {https://ui.adsabs.harvard.edu/abs/2021ApJ...916...77P} {916, 77}

\bibitem[\protect\citeauthoryear{{Quirrenbach} et~al.,}{{Quirrenbach} et~al.}{2010}]{Quirrenbach_2010}
{Quirrenbach} A.,  et~al., 2010, in {McLean} I.~S.,  {Ramsay} S.~K.,   {Takami} H.,  eds,  Society of Photo-Optical Instrumentation Engineers (SPIE) Conference Series Vol. 7735, Ground-based and Airborne Instrumentation for Astronomy III. p. 773513, \mn@doi{10.1117/12.857777}

\bibitem[\protect\citeauthoryear{{Rackham}, {Apai}  \& {Giampapa}}{{Rackham} et~al.}{2018}]{Rackham2018}
{Rackham} B.~V.,  {Apai} D.,   {Giampapa} M.~S.,  2018, \mn@doi [\apj] {10.3847/1538-4357/aaa08c}, \href {https://ui.adsabs.harvard.edu/abs/2018ApJ...853..122R} {853, 122}

\bibitem[\protect\citeauthoryear{{Rackham} et~al.,}{{Rackham} et~al.}{2023}]{Rackham2023}
{Rackham} B.~V.,  et~al., 2023, \mn@doi [RAS Techniques and Instruments] {10.1093/rasti/rzad009}, \href {https://ui.adsabs.harvard.edu/abs/2023RASTI...2..148R} {2, 148}

\bibitem[\protect\citeauthoryear{Rasmussen \& Williams}{Rasmussen \& Williams}{2006}]{Rasmussen2006}
Rasmussen C.~E.,  Williams C. K.~I.,  2006, {Gaussian Processes for Machine Learning}.
MIT Press, Cambridge, MA, \url {https://ui.adsabs.harvard.edu/abs/2006gpml.book.....R}

\bibitem[\protect\citeauthoryear{{Reyl{\'e}}, {Jardine}, {Fouqu{\'e}}, {Caballero}, {Smart}  \& {Sozzetti}}{{Reyl{\'e}} et~al.}{2021}]{Reyle_2021}
{Reyl{\'e}} C.,  {Jardine} K.,  {Fouqu{\'e}} P.,  {Caballero} J.~A.,  {Smart} R.~L.,   {Sozzetti} A.,  2021, \mn@doi [\aap] {10.1051/0004-6361/202140985}, \href {https://ui.adsabs.harvard.edu/abs/2021A&A...650A.201R} {650, A201}

\bibitem[\protect\citeauthoryear{{Ribas}}{{Ribas}}{2007}]{Ribas2007}
{Ribas} I.,  2007, \mn@doi [Highlights of Astronomy] {10.1017/S1743921307010678}, \href {https://ui.adsabs.harvard.edu/abs/2007HiA....14..295R} {14, 295}

\bibitem[\protect\citeauthoryear{{Ribas} et~al.,}{{Ribas} et~al.}{2023}]{Ribas_2023}
{Ribas} I.,  et~al., 2023, \mn@doi [\aap] {10.1051/0004-6361/202244879}, \href {https://ui.adsabs.harvard.edu/abs/2023A&A...670A.139R} {670, A139}

\bibitem[\protect\citeauthoryear{{Rigby} et~al.,}{{Rigby} et~al.}{2024}]{Rigby_towards}
{Rigby} F.~E.,  et~al., 2024, \mn@doi [arXiv e-prints] {10.48550/arXiv.2409.03683}, \href {https://ui.adsabs.harvard.edu/abs/2024arXiv240903683R} {p. arXiv:2409.03683}

\bibitem[\protect\citeauthoryear{{Robinson}, {Cram}  \& {Giampapa}}{{Robinson} et~al.}{1990}]{Robinson1990}
{Robinson} R.~D.,  {Cram} L.~E.,   {Giampapa} M.~S.,  1990, \mn@doi [\apjs] {10.1086/191525}, \href {https://ui.adsabs.harvard.edu/abs/1990ApJS...74..891R} {74, 891}

\bibitem[\protect\citeauthoryear{{Rodr{\'\i}guez-Mozos} \& {Moya}}{{Rodr{\'\i}guez-Mozos} \& {Moya}}{2019}]{RMM2019}
{Rodr{\'\i}guez-Mozos} J.~M.,  {Moya} A.,  2019, \mn@doi [\aap] {10.1051/0004-6361/201935543}, \href {https://ui.adsabs.harvard.edu/abs/2019A&A...630A..52R} {630, A52}

\bibitem[\protect\citeauthoryear{{Ros{\'e}n}, {Kochukhov}  \& {Wade}}{{Ros{\'e}n} et~al.}{2015}]{Rosen2015}
{Ros{\'e}n} L.,  {Kochukhov} O.,   {Wade} G.~A.,  2015, \mn@doi [\apj] {10.1088/0004-637X/805/2/169}, \href {https://ui.adsabs.harvard.edu/abs/2015ApJ...805..169R} {805, 169}

\bibitem[\protect\citeauthoryear{Sairam \& Triaud}{Sairam \& Triaud}{2022}]{Lalitha_2022}
Sairam L.,  Triaud A. H. M.~J.,  2022, \mn@doi [Monthly Notices of the Royal Astronomical Society] {10.1093/mnras/stac1446}, 514, 2259

\bibitem[\protect\citeauthoryear{{Santos}, {Gomes da Silva}, {Lovis}  \& {Melo}}{{Santos} et~al.}{2010}]{Santos2010}
{Santos} N.~C.,  {Gomes da Silva} J.,  {Lovis} C.,   {Melo} C.,  2010, \mn@doi [\aap] {10.1051/0004-6361/200913433}, \href {https://ui.adsabs.harvard.edu/abs/2010A&A...511A..54S} {511, A54}

\bibitem[\protect\citeauthoryear{{Sanz-Forcada}, {Micela}, {Ribas}, {Pollock}, {Eiroa}, {Velasco}, {Solano}  \& {Garc{\'\i}a-{\'A}lvarez}}{{Sanz-Forcada} et~al.}{2011}]{Sanz_2011}
{Sanz-Forcada} J.,  {Micela} G.,  {Ribas} I.,  {Pollock} A.~M.~T.,  {Eiroa} C.,  {Velasco} A.,  {Solano} E.,   {Garc{\'\i}a-{\'A}lvarez} D.,  2011, \mn@doi [\aap] {10.1051/0004-6361/201116594}, \href {https://ui.adsabs.harvard.edu/abs/2011A&A...532A...6S} {532, A6}

\bibitem[\protect\citeauthoryear{{Sarkis} et~al.,}{{Sarkis} et~al.}{2018}]{Sarkis_2018}
{Sarkis} P.,  et~al., 2018, \mn@doi [\aj] {10.3847/1538-3881/aac108}, \href {https://ui.adsabs.harvard.edu/abs/2018AJ....155..257S} {155, 257}

\bibitem[\protect\citeauthoryear{{Schwarz}}{{Schwarz}}{1978}]{Schwarz1978}
{Schwarz} G.,  1978, Annals of Statistics, \href {https://ui.adsabs.harvard.edu/abs/1978AnSta...6..461S} {6, 461}

\bibitem[\protect\citeauthoryear{{Semel}}{{Semel}}{1989}]{Semel1989}
{Semel} M.,  1989, \aap, \href {https://ui.adsabs.harvard.edu/abs/1989A&A...225..456S} {225, 456}

\bibitem[\protect\citeauthoryear{{Shan} et~al.,}{{Shan} et~al.}{2024}]{Shan2024}
{Shan} Y.,  et~al., 2024, \mn@doi [\aap] {10.1051/0004-6361/202346794}, \href {https://ui.adsabs.harvard.edu/abs/2024A&A...684A...9S} {684, A9}

\bibitem[\protect\citeauthoryear{Shields, Ballard  \& Johnson}{Shields et~al.}{2016}]{SHIELDS20161}
Shields A.~L.,  Ballard S.,   Johnson J.~A.,  2016, \mn@doi [Physics Reports] {https://doi.org/10.1016/j.physrep.2016.10.003}, 663, 1

\bibitem[\protect\citeauthoryear{{Shulyak} et~al.,}{{Shulyak} et~al.}{2019}]{Shulyak2019}
{Shulyak} D.,  et~al., 2019, \mn@doi [\aap] {10.1051/0004-6361/201935315}, \href {https://ui.adsabs.harvard.edu/abs/2019A&A...626A..86S} {626, A86}

\bibitem[\protect\citeauthoryear{{Sissa} et~al.,}{{Sissa} et~al.}{2016}]{Sissa2016}
{Sissa} E.,  et~al., 2016, \mn@doi [\aap] {10.1051/0004-6361/201628531}, \href {https://ui.adsabs.harvard.edu/abs/2016A&A...596A..76S} {596, A76}

\bibitem[\protect\citeauthoryear{{Strassmeier}, {Fekel}, {Bopp}, {Dempsey}  \& {Henry}}{{Strassmeier} et~al.}{1990}]{Strassmeier1990}
{Strassmeier} K.~G.,  {Fekel} F.~C.,  {Bopp} B.~W.,  {Dempsey} R.~C.,   {Henry} G.~W.,  1990, \mn@doi [\apjs] {10.1086/191414}, \href {https://ui.adsabs.harvard.edu/abs/1990ApJS...72..191S} {72, 191}

\bibitem[\protect\citeauthoryear{Strassmeier et~al.,}{Strassmeier et~al.}{2004}]{Strassmeier_2004}
Strassmeier K.~G.,  et~al., 2004, \mn@doi [Astronomische Nachrichten] {https://doi.org/10.1002/asna.200410273}, 325, 527

\bibitem[\protect\citeauthoryear{{Su{\'a}rez Mascare{\~n}o} et~al.,}{{Su{\'a}rez Mascare{\~n}o} et~al.}{2018}]{Suarez_Mascareno_2018}
{Su{\'a}rez Mascare{\~n}o} A.,  et~al., 2018, \mn@doi [\aap] {10.1051/0004-6361/201732143}, \href {https://ui.adsabs.harvard.edu/abs/2018A&A...612A..89S} {612, A89}

\bibitem[\protect\citeauthoryear{Thompson et~al.,}{Thompson et~al.}{2024}]{Thompson_2024}
Thompson A.,  et~al., 2024, \mn@doi [The Astrophysical Journal] {10.3847/1538-4357/ad0369}, 960, 107

\bibitem[\protect\citeauthoryear{{Tsai}, {Innes}, {Lichtenberg}, {Taylor}, {Malik}, {Chubb}  \& {Pierrehumbert}}{{Tsai} et~al.}{2021}]{Tsai2021}
{Tsai} S.-M.,  {Innes} H.,  {Lichtenberg} T.,  {Taylor} J.,  {Malik} M.,  {Chubb} K.,   {Pierrehumbert} R.,  2021, \mn@doi [\apjl] {10.3847/2041-8213/ac399a}, \href {https://ui.adsabs.harvard.edu/abs/2021ApJ...922L..27T} {922, L27}

\bibitem[\protect\citeauthoryear{{Tsiaras}, {Waldmann}, {Tinetti}, {Tennyson}  \& {Yurchenko}}{{Tsiaras} et~al.}{2019}]{Tsiaras2019}
{Tsiaras} A.,  {Waldmann} I.~P.,  {Tinetti} G.,  {Tennyson} J.,   {Yurchenko} S.~N.,  2019, Nature Astronomy, p.~451

\bibitem[\protect\citeauthoryear{{Valencia}, {Ikoma}, {Guillot}  \& {Nettelmann}}{{Valencia} et~al.}{2010}]{Valencia2010}
{Valencia} D.,  {Ikoma} M.,  {Guillot} T.,   {Nettelmann} N.,  2010, \mn@doi [\aap] {10.1051/0004-6361/200912839}, \href {https://ui.adsabs.harvard.edu/abs/2010A&A...516A..20V} {516, A20}

\bibitem[\protect\citeauthoryear{{Vida} \& {Roettenbacher}}{{Vida} \& {Roettenbacher}}{2018}]{Vida2018}
{Vida} K.,  {Roettenbacher} R.~M.,  2018, \mn@doi [\aap] {10.1051/0004-6361/201833194}, \href {https://ui.adsabs.harvard.edu/abs/2018A&A...616A.163V} {616, A163}

\bibitem[\protect\citeauthoryear{Vidotto et~al.,}{Vidotto et~al.}{2014}]{Vidotto_2014}
Vidotto A.~A.,  et~al., 2014, \mn@doi [Monthly Notices of the Royal Astronomical Society] {10.1093/mnras/stu728}, 441, 2361

\bibitem[\protect\citeauthoryear{{Vogt}, {Penrod}  \& {Hatzes}}{{Vogt} et~al.}{1987}]{Vogt1987}
{Vogt} S.~S.,  {Penrod} G.~D.,   {Hatzes} A.~P.,  1987, \mn@doi [\apj] {10.1086/165647}, \href {https://ui.adsabs.harvard.edu/abs/1987ApJ...321..496V} {321, 496}

\bibitem[\protect\citeauthoryear{{Watson}, {Donahue}  \& {Walker}}{{Watson} et~al.}{1981}]{Watson_1981}
{Watson} A.~J.,  {Donahue} T.~M.,   {Walker} J.~C.~G.,  1981, \mn@doi [\icarus] {10.1016/0019-1035(81)90101-9}, \href {https://ui.adsabs.harvard.edu/abs/1981Icar...48..150W} {48, 150}

\bibitem[\protect\citeauthoryear{{Weisskopf}, {Tananbaum}, {Van Speybroeck}  \& {O'Dell}}{{Weisskopf} et~al.}{2000}]{Weisskopf2000}
{Weisskopf} M.~C.,  {Tananbaum} H.~D.,  {Van Speybroeck} L.~P.,   {O'Dell} S.~L.,  2000, in {Truemper} J.~E.,  {Aschenbach} B.,  eds,  Society of Photo-Optical Instrumentation Engineers (SPIE) Conference Series Vol. 4012, X-Ray Optics, Instruments, and Missions III. pp 2--16 (\mn@eprint {arXiv} {astro-ph/0004127}), \mn@doi{10.1117/12.391545}

\bibitem[\protect\citeauthoryear{{Weisskopf}, {Brinkman}, {Canizares}, {Garmire}, {Murray}  \& {Van Speybroeck}}{{Weisskopf} et~al.}{2002}]{Weisskopf2002}
{Weisskopf} M.~C.,  {Brinkman} B.,  {Canizares} C.,  {Garmire} G.,  {Murray} S.,   {Van Speybroeck} L.~P.,  2002, \mn@doi [\pasp] {10.1086/338108}, \href {https://ui.adsabs.harvard.edu/abs/2002PASP..114....1W} {114, 1}

\bibitem[\protect\citeauthoryear{Wise, Dodson-Robinson, Bevenour  \& Provini}{Wise et~al.}{2018}]{Wise_2018}
Wise A.~W.,  Dodson-Robinson S.~E.,  Bevenour K.,   Provini A.,  2018, \mn@doi [The Astronomical Journal] {10.3847/1538-3881/aadd94}, 156, 180

\bibitem[\protect\citeauthoryear{Worthey \& chul Lee}{Worthey \& chul Lee}{2011}]{Worthey_2011}
Worthey G.,  chul Lee H.,  2011, \mn@doi [The Astrophysical Journal Supplement Series] {10.1088/0067-0049/193/1/1}, 193, 1

\bibitem[\protect\citeauthoryear{{Yuk} et~al.,}{{Yuk} et~al.}{2010}]{Yuk_2010}
{Yuk} I.-S.,  et~al., 2010, in {McLean} I.~S.,  {Ramsay} S.~K.,   {Takami} H.,  eds,  Society of Photo-Optical Instrumentation Engineers (SPIE) Conference Series Vol. 7735, Ground-based and Airborne Instrumentation for Astronomy III. p. 77351M, \mn@doi{10.1117/12.856864}

\bibitem[\protect\citeauthoryear{{Zacharias}, {Finch}, {Girard}, {Henden}, {Bartlett}, {Monet}  \& {Zacharias}}{{Zacharias} et~al.}{2012}]{Zacharias2012}
{Zacharias} N.,  {Finch} C.~T.,  {Girard} T.~M.,  {Henden} A.,  {Bartlett} J.~L.,  {Monet} D.~G.,   {Zacharias} M.~I.,  2012, {VizieR Online Data Catalog: UCAC4 Catalogue (Zacharias+, 2012)}, VizieR On-line Data Catalog: I/322A. Originally published in: 2013AJ....145...44Z

\bibitem[\protect\citeauthoryear{do Amaral, Barnes, Segura  \& Luger}{do~Amaral et~al.}{2022}]{Amaral_2022}
do Amaral L. N.~R.,  Barnes R.,  Segura A.,   Luger R.,  2022, \mn@doi [The Astrophysical Journal] {10.3847/1538-4357/ac53af}, 928, 12

\bibitem[\protect\citeauthoryear{{dos Santos} et~al.,}{{dos Santos} et~al.}{2020}]{Santos_2020}
{dos Santos} L.~A.,  et~al., 2020, \mn@doi [\aap] {10.1051/0004-6361/201937327}, \href {https://ui.adsabs.harvard.edu/abs/2020A&A...634L...4D} {634, L4}

\bibitem[\protect\citeauthoryear{Ítalo G.~Gonçalves, Echer  \& Frigo}{Ítalo G.~Gonçalves et~al.}{2020}]{GONCALVES2020677}
Ítalo G.~Gonçalves Echer E.,   Frigo E.,  2020, \mn@doi [Advances in Space Research] {https://doi.org/10.1016/j.asr.2019.11.011}, 65, 677

\makeatother
\end{thebibliography}


\begin{appendix}

\section{Stellar rotation period}

In Table~\ref{tab:prior}, we list the priors used for the GP model 
paramaters in our analysis in section \S\ref{sec:rot_k2_18} and \ref{sec:rot_toi732}. The mean parameter represents the baseline flux around which the variation in the stellar light curve occurs and the jitter term accounts for the excess noise beyond the uncertainities reported in the flux measurements. The parameters Sigma, rho and Sigma$_{\mathrm{rot}}$ represents the overall variability, the characteristic timescale for correlations and the amplitude of the rotational term, respectively. The term ln period refers to the logarithm of stellar rotation period. Q0 and dQ are the  quality factors, with Q0 representing the sharpness of the rotational periodic signal and dQ controlling the damping of periodic component over time. The choice of these priors follows the approach outlined in \citet{exoplanet:joss}.
\begin{table}
\caption{Priors for the GP model parameters}
\begin{tabular}{|l|c|}
\hline
\textbf{Parameter}  & \textbf{Prior}  \\
\hline
Mean                & $\mathcal{N}(0, 10)$ \\
ln jitter         & $\mathcal{N}(log \langle \mathrm{Flux_{err}} \rangle, 2)$ \\
Sigma               & $\mathcal{IG}(1.0, 5)$ \\
Rho                 & $\mathcal{IG}(0.5, 2)$ \\
Sigma$_{\mathrm{rot}}$          & $\mathcal{IG}(1, 5)$ \\
ln period         & $\mathcal{N}(log\langle \mathrm{peak_{period}}\rangle, 2)$ \\
ln Q0             & $\mathcal{HN}(0,2)$
 \\
ln dQ             & $\mathcal{N}(0,2)$
 \\
\hline
\end{tabular}

Note: Normal distribution is represented as $\mathcal{N}$($\mu$, $\sigma$) where $\mu$ is the mean and $\sigma$ is the standard deviation; Inverse-Gamma distribution is  represented as $\mathcal{IG}$($\alpha$, $\beta$) where $\alpha$ is the shape parameter and $\beta$ is the scale parameter; Half-Normal distribution is written as $\mathcal{HN}$(0, $\sigma$) where $\sigma$ is the standard deviation. Peak$_{\mathrm{period}}$ is the period from the Lomb-scargle periodogram.
\label{tab:prior}
\end{table}

\begin{table}
    \centering
        \caption{Best fitting rotation periods (P$_{\text{rot}})$, with their uncertainties.}
    \begin{tabular}{cllcc}
        \hline
        \textbf{Mode} & \textbf{Filter} & \textbf{Period [days]} \\
        \hline
        & K2-18&\\
        Photometry & Stella B filter & 39.2 $^{+2.1}_{-1.8}$\\
                   & Stella R filter & 38.9 $^{+2.3}_{-1.8}$\\
                   & ASAS  V  filter  & 38.9 $^{+2.1}_{-1.7}$\\
                   & Kepler          & 39.8 $^{+3.4}_{-0.4}$\\
                   & CATALINA & 38.9 $^{+1.0}_{-3.2}$\\
      Spectroscopy & HARPS H-alpha   & 40.2 $^{+3.9}_{-0.8}$\\
                   & HARPS Na D      & 43.8 $^{+0.4}_{-4.1}$\\
                   & HARPS He I      & 39.7 $^{+2.3}_{-0.8}$\\
                   & HARPS Ca I      & 39.5 $^{+3.0}_{-2.9}$\\
                   & CARMENES H-alpha & 33.3 $^{+5.1}_{-1.8}$ \\
                   & CARMENES Na D   & 30.1 $^{+6.8}_{-5.3}$ \\
        \hline
        & TOI-732&\\
        Photometry & ASAS V filter & 143.6$^{+7.8}_{-6.6}$\\
                   & M Earth  & 136.4$^{+9.4}_{-4.2}$\\
                   & CATALINA & 135.6$^{+6.9}_{-6.3}$\\
      Spectroscopy & HARPS H-alpha   & 105.7 $^{+9.5}_{-3.4}$\\
                   & HARPS Na D      & 62.7 $^{+12.8}_{-1.1}$\\
                   & HARPS He I      & 81.8 $^{+15.4}_{-1.8}$\\
                   & HARPS Ca I      & 58.0 $^{+13.5}_{-0.8}$\\

        \hline
    \end{tabular}

    \label{tab:rot_periods}
\end{table}

\section{TOI 732 - rotation period and equivalent widths}

To investigate the nature and variability of TOI-732, we investigate the behaviour of several key spectral lines. 

\begin{figure*}
    \centering
    \includegraphics[width=0.49\textwidth]{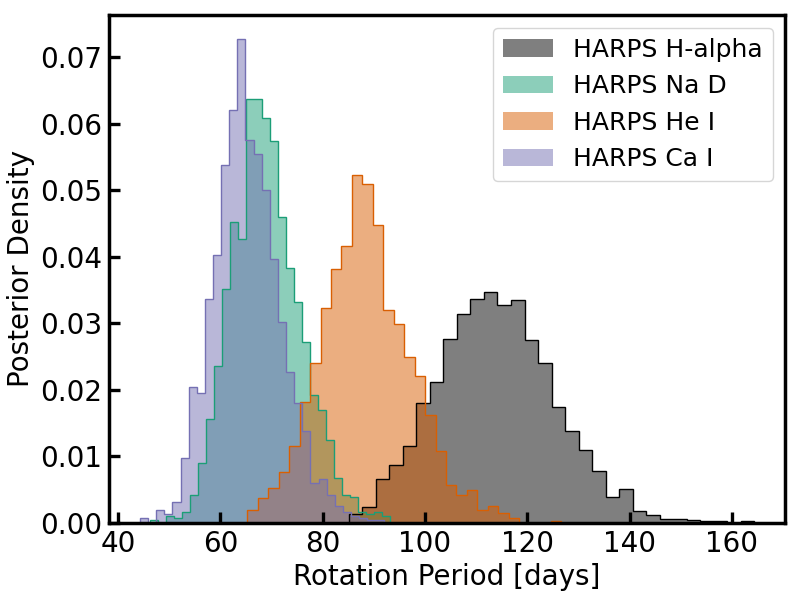}
    \includegraphics[width=0.43\textwidth]{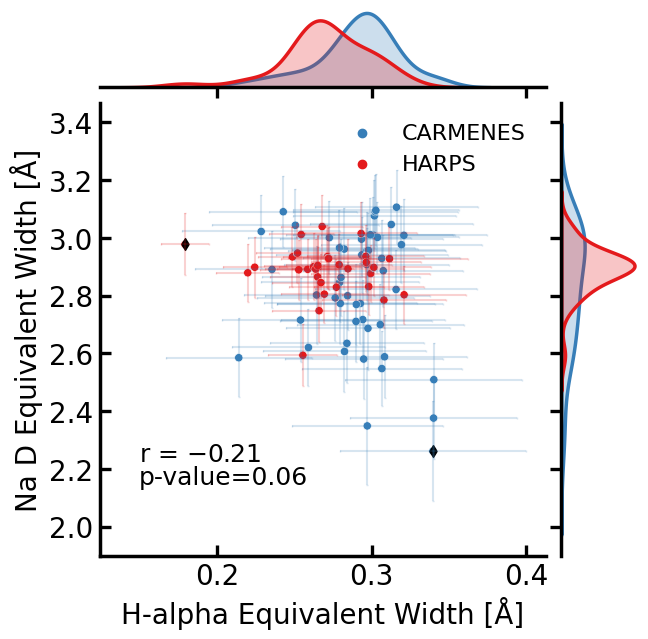}
    \caption{Left panel:  Posterior distribution for activity indicators
observed with the HARPS spectrograph for TOI-732, including H-alpha (grey), Na D
(green), He I (orange), and Ca I (purple). Right panel: Correlation plot between H-alpha and NaD equivalent widths with error bars indicating uncertainties measured with HARPS for TOI-732. Colours distinguish outliers identified using a z-score threshold of 3 (black diamonds). This weak negative correlation suggests minimal influence of one line on the other.}
    \label{fig:toi732_correlation}
\end{figure*}

\begin{figure*}
    \centering
    \includegraphics[width=\textwidth]{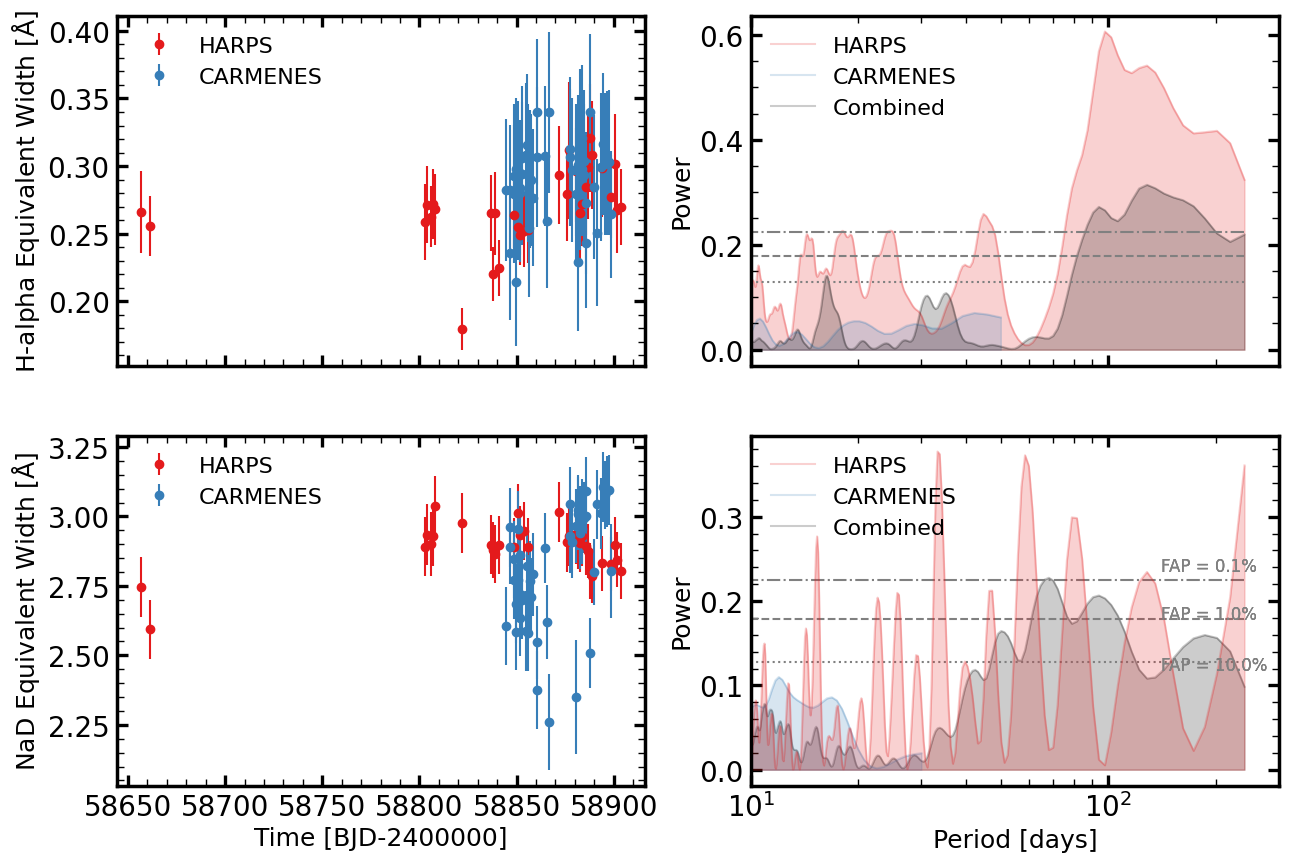}
    \caption{The EqW time series and Lomb-Scargle periodograms for H-alpha (top panels) and NaD (bottom panels) lines measured with the HARPS and CARMENES for TOI-732.}
    \label{fig:toi732_ts}
\end{figure*}

\begin{figure*}
    \centering
    \includegraphics[width=0.49\textwidth]{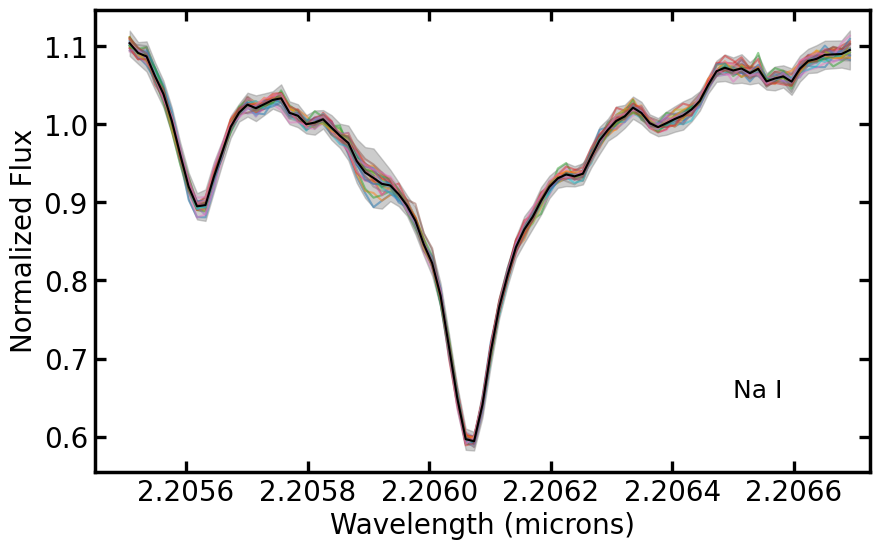}
    \includegraphics[width=0.49\textwidth]{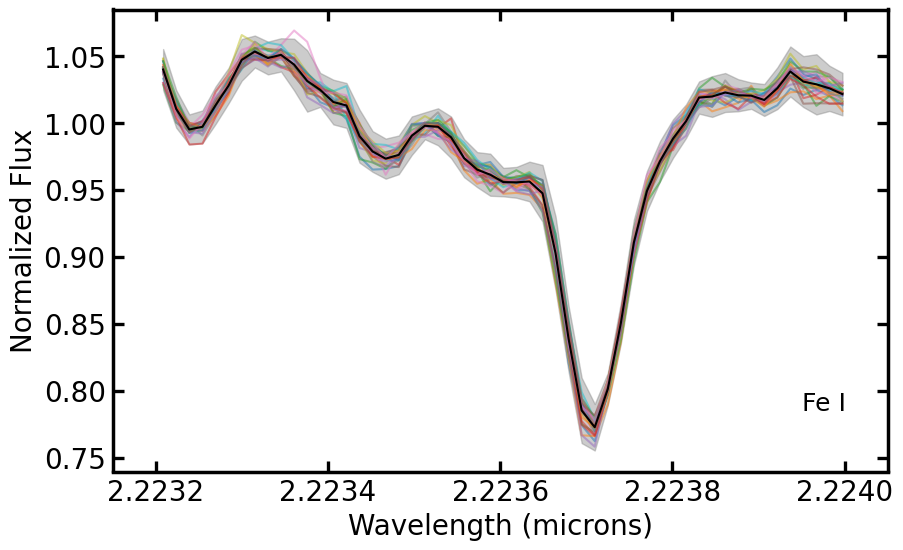}
    \caption{IGRINS spectra of Na I (left) and Fe I (right) lines for TOI-732. The mean profile is shown in black, and gray areas represent the variance.}
    \label{fig:igrins_spec_line}
\end{figure*}

Figures \ref{fig:toi732_correlation} (right panel), \ref{fig:toi732_ts}, and \ref{fig:igrins_spec_line} present a detailed analysis of H-alpha and NaD at optical wavelengths and Na I and Fe I at infrared wavelengths. By studying correlations, time-series variations, and line profiles, we aim to characterise the stellar activity and understand its potential impact on exoplanet observations.

\section{Activity cycle predictions}

\begin{figure*}
\includegraphics[width=0.99\textwidth]{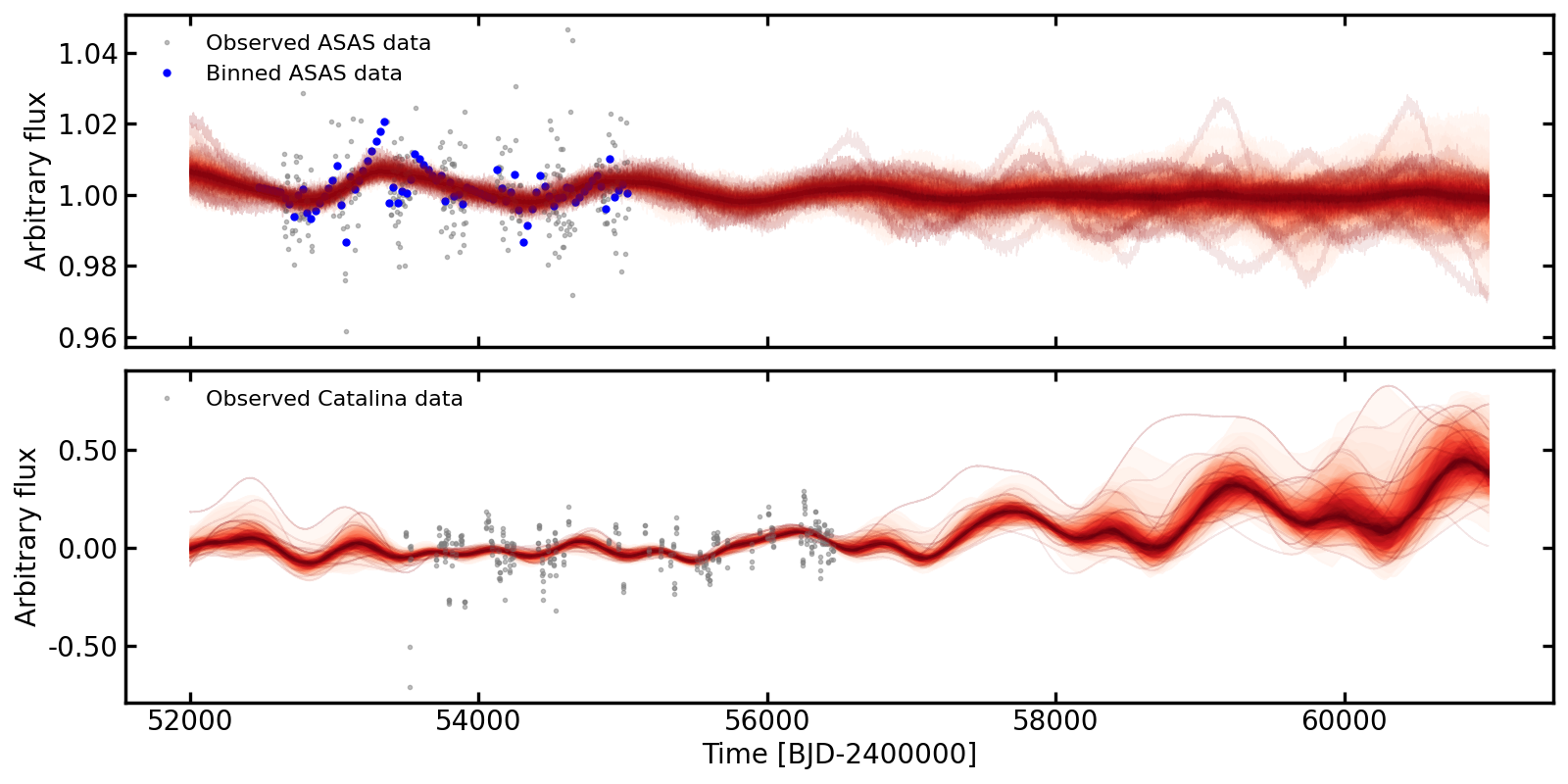}
\caption{ASAS V-band (top panel) and CATALINA (bottom panel) light curves of K2-18. Blue points represent binned data, red curve shows the posterior distribution of GP model prediction of activity cycle.}
\label{fig:staccato_k2_18}
\end{figure*}

Figure \ref{fig:staccato_k2_18} shows GP models fitted to ASAS V-band and CATALINA light curves of K2-18. By modelling the star's variability, we aim to predict its activity cycle phase and assess the potential impact on future exoplanet atmospheric observations. This information can be helpful for optimising observation planning and minimising the influence of stellar activity on data interpretation.

\end{appendix}

\end{document}